\documentclass[a4paper,preprint]{elsarticle}
\usepackage[utf8]{inputenc}
\usepackage{graphics}
\usepackage[dvipsnames]{xcolor}
\usepackage{amsmath}
\usepackage{amsfonts}
\usepackage{hyperref}
\usepackage{cancel}
\usepackage{graphicx}
\usepackage{amssymb}
\usepackage{chemformula}
\usepackage{mathtools} 

\newcommand{\avec}{\boldsymbol{a}}

\newcommand{\nvec}{\boldsymbol{n}}
\newcommand{\qvec}{\boldsymbol{q}}
\newcommand{\kvec}{\boldsymbol{k}}

\newcommand{\Pvec}{\boldsymbol{P}}

\begin{document}

\begin{frontmatter}
\title{Fermion pairing in body-centered-cubic quantum simulators of extended Hubbard models}

\author{Ganiyu D. Adebanjo}
\address{School of Physical Sciences, The Open University, Walton Hall, Milton Keynes, MK7 6AA, UK}

\author{P.E. Kornilovitch}
\address{Department of Physics, Oregon State University, Corvallis, OR, 97331, USA}

\author{J.P. Hague}
\address{School of Physical Sciences, The Open University, Walton Hall, Milton Keynes, MK7 6AA, UK}

\date{\today}

\begin{abstract}
We investigate formation and condensation of fermion pairs in cold-atom quantum simulators for extended Hubbard models ($UV$ models) with body-centered-cubic (BCC) optical lattices in the dilute limit, predicting small and light pairs. Pair mass, radius, and binding conditions are calculated, and used to compute transition temperatures. We predict that: (a) local pairs form in BCC optical lattices and binding energies can be large; (b) for particular cases where onsite $U$ and intersite $V$ are attractive with similar size, pairs are both small and light; and (c) pairs of $^6$Li atoms Bose--Einstein condense at temperatures of around 10 nK.
\end{abstract}

\end{frontmatter}


\section{Introduction}

 Optical lattices with BCC structures can be formed using arrays of four laser beams \cite{yuan2003}, and  are of interest for two reasons. Firstly, they have been largely neglected in the context of quantum simulators. Secondly, there are condensed matter systems of interest with BCC lattices that could benefit from the insight provided by quantum simulators, such as BCC A$_3$C$_{60}$ superconductors, which have high transition temperatures (38K)\cite{ganinetal2008}. The goal of this article is to discuss the properties of fermion pairs formed by extended Hubbard interactions in cold-atom quantum simulators with BCC lattices.

The ability to probe Hubbard models in clean and well-controlled systems \cite{bloch2008} has been a major success of cold atom  quantum simulators  formed using optical lattices. Quantum simulators offer the possibility to implement Hubbard models, in a way that cannot be achieved in condensed matter. For example single-band Hubbard models can be implemented without the complications of interactions between multiple electronic bands  \cite{bloch2008}.  Several milestones have been achieved using cold atoms in optical lattices, including observations of Mott transitions in repulsive Hubbard models \cite{greiner2002,jordens2008}. The interactions in cold-atom quantum simulators can be tuned such that attractive Hubbard models can be studied, allowing local pairs to be observed \cite{mitra2018,brown2020}.

A simple extension to the Hubbard model\cite{hubbard1963electron}, known as the extended Hubbard model\cite{hirsch1984}, or (in the low density limit) $UV$-model\cite{kornilovitch2004}, includes an onsite Hubbard $U$ and an intersite interaction $V$. 
The $UV$ Hamiltonian is defined as:
\begin{equation}\label{themodelHamiltonian}
H = \sum_{\langle\nvec,\avec\rangle\sigma}t_{\avec}\,c_{\nvec+\avec,\sigma}^{\dagger}\,c_{\nvec\sigma} + U\sum_{\nvec} \hat{\rho}_{\nvec \uparrow}\,\hat{\rho}_{\nvec \downarrow} +   \sum_{\langle\nvec,\avec\rangle}V\:\hat{\rho}_{\nvec + \avec}\:\hat{\rho}_{\nvec}
\end{equation}
where $c^{\dagger}_{\nvec\sigma}$ ($c_{\nvec\sigma}$) creates (annihilates) an atom of spin $\sigma$ at site $\nvec$, $\hat{\rho}_{\nvec}=\hat{\rho}_{\nvec\uparrow}+\hat{\rho}_{\nvec\downarrow}$, where $\hat{\rho}_{\nvec\sigma}$ is the number operator for atoms on site $\nvec$ with spin $\sigma$, $\avec$ the intersite lattice vector, $t_{\avec}$ is the intersite hopping, $U$ is the onsite interaction and $V$ is the intersite interaction. Both $U$ and $V$ may be attractive or repulsive. For a BCC lattice, $|\avec|=\frac{\sqrt{3}}{2}b$ where $b$ is the lattice constant. 

$UV$ models are of interest because local Coulomb repulsion and an intersite effective attraction are key features of many unconventional superconductors\cite{micnas1990}. The site-local Hubbard $U$ is typically present in any superconductor with low kinetic energy. Since Coulomb repulsion is typically small between sites due to screening (especially in 3D) an effective intersite attraction or repulsion could arise due to phonons, spin fluctuations or other exotic pairing mechanisms\cite{micnas1990}. For example, an effective $UV$ model can be generated in systems with long-range electron-phonon coupling and local Coulomb repulsion (Hubbard $U$).

In optical lattices, very large $U$ and $V$ with both attractive and repulsive signs can be achieved relative to those in condensed matter systems. $U$ can be changed by orders of magnitude by modifying the magnetic field at the Feshbach resonance\cite{bloch2008}. Intersite $V$ can be achieved using dressed Rydberg states, which are highly polarizable, and have sizable long range interactions \cite{pupillo2009,hague2014}. There are several ways in which $V$ can be tuned. The simplest is by modifying the principal quantum number of the Rydberg state. The dipole-dipole (van der Walls) interaction between Rydberg states increases with the fourth (eleventh) power of the principal quantum number $n$, leading to a high level of control over the size of long range interactions\cite{sibalic2018}. Interactions between dressed Rydberg states can be highly isotropic\footnote{There is a high level of control as to whether interactions are isotropic or anisotropic via the Forster resonance.} and thus suitable for generating an intersite $V$ \cite{walker2008} in a quantum simulator. By combining dressed Rydberg states with Feshbach resonances, a $UV$ model can be realized on an optical lattice\cite{hague2014}. 

In this paper we investigate the properties of local pairs formed from extended Hubbard interactions in BCC optical lattices, which to our knowledge have not been studied in this context. We calculate the critical pair attraction strength $U_{c}$ ($V_{c}$) required for the formation of onsite (intersite) pairs. Pair dispersion, mass and size are determined. We estimate the BEC transition temperature for local fermion pairs. The paper is organized as follows: We describe the methodology used to solve the $UV$ model in the low density limit (Sec. \ref{sec:methodology}). In Sec. \ref{sec:pair_properties}, we report the properties of local pairs formed in BCC lattices. We conclude this work with discussion in Sec. \ref{sec:discussion}.

\section{Methodology}
\label{sec:methodology}

In this section, we describe the steps required to obtain solutions to the $UV$ model with low particle density (dilute limit). We study the $UV$ model with two spin-$\frac{1}{2}$ fermions.
We solve Hamiltonian (\ref{themodelHamiltonian}) by constructing a real-space Schr\"{o}dinger equation. The two-body wave function $\Psi({\nvec_{1}}\,{\nvec_{2}})$ must satisfy the equation:
\begin{align}
& \sum_{\avec}t_{\avec}\,[\Psi(\nvec_{1}+\avec,\nvec_{2})\nonumber\\
& +\Psi(\nvec_{1},\nvec_{2}+\avec)] + \sum_{\avec}\hat{V}_{\avec}\,\delta_{\nvec_{1} - \nvec_{2},\avec}\Psi(\nvec_{1},\nvec_{2}) \nonumber\\
& = E\,\Psi(\nvec_{1},\nvec_{2}) 
\label{twobody wavefunction S.E}
\end{align}
\noindent where the interaction terms have been combined into a single function (i.e. $\hat{V}_{\avec=0}=U$ and $\hat{V}_{\avec \ne0}=V$), $E$ is the total energy of the system.

Equation (\ref{twobody wavefunction S.E}) can be solved as follows. We construct the wave function in momentum space using a Fourier transform
\begin{equation}\label{FT wave function}
\psi_{\kvec_{1}\kvec_{2}} = \frac{1}{N}\sum_{\nvec_{1}\nvec_{2}}\Psi(\nvec_{1},\nvec_{2})\,e^{-i\kvec_{1}\,\nvec_{1} - i\kvec_{2}\,\nvec_{2}}
\end{equation}
\noindent with $N$ being the total number of lattice points.

Then, substituting Equation (\ref{FT wave function}) into Equation (\ref{twobody wavefunction S.E}), we obtain:
\begin{equation}\label{equation_four}
(E-\varepsilon_{\kvec_{1}}-\varepsilon_{\kvec_{2}})\psi_{\kvec_{1}\kvec_{2}}=\frac{1}{N}\sum_{\avec\qvec}\hat{V}_{\avec}\,e^{i(\qvec -\kvec_{1})\avec}\,\psi_{\qvec,\kvec_{1}+\kvec_{2}-\qvec} \;\;\;,
\end{equation}
with 
\begin{equation}\label{one_p_disp_relation}
\varepsilon_{\kvec}=\sum_{\avec}t_{\avec}\,e^{i\kvec\cdot \avec}
\end{equation}
\noindent
being the one-particle energy dispersion of the model, where $\kvec$ is the particle's momentum vector.

We define
\begin{equation}\label{equation_six}
\Phi_{\avec}(\kvec_{1}+\kvec_{2})=\Phi_{\avec}(\Pvec)\equiv \frac{1}{N}\sum_{\qvec}e^{i\qvec\avec}\;\psi_{\qvec,\Pvec-\qvec}
\end{equation}
where $\Pvec=\kvec_{1}+\kvec_{2}$ is the total momentum of the particle pair. By using this definition, Equation (\ref{equation_four}) can then be rewritten as

\begin{equation}\label{equation_seven}
\psi_{\kvec_{1}\kvec_{2}}=\sum_{\avec}\hat{V}_{\avec}\,\frac{e^{-i\kvec_{1}\avec}}{E-\varepsilon_{\kvec_{1}}-\varepsilon_{\kvec_{2}}}\Phi_{\avec}(\Pvec)
\end{equation}

\noindent and finally this expression is substituted into Equation (\ref{equation_six}) which becomes

\begin{equation}\label{equation_eight}
\Phi_{\avec}(\Pvec)=-\sum_{\avec\prime}\hat{V}_{\avec^\prime}\;L_{\avec\avec^\prime}(E,\Pvec)\;\Phi_{\avec^\prime}(\Pvec)
\end{equation}
where the Green's function, $L_{\avec\avec^\prime}(E,\Pvec)$, is determined from the following sum,
\begin{equation}\label{equation_nine}
L_{\avec\avec^\prime}(E,\Pvec)=\frac{1}{N}\sum_{\qvec}\frac{e^{i\qvec\,(\avec-\avec^\prime)}}{-E+\varepsilon_{\qvec}+\varepsilon_{\Pvec-\qvec}}.
\end{equation}

The set of linear equations (\ref{equation_eight}) are solved by 
\begin{equation}\label{equation_ten}
\mathrm{det}\mid -\hat{V}_{\avec^\prime}\;L_{\avec\avec^\prime}(E,\Pvec)-\delta_{\avec\avec^\prime} \mid = 0 ,
\end{equation} 
which determines the system energy $E(\Pvec)$ as a function of the total momentum of the pair. Equations (\ref{equation_seven})--(\ref{equation_ten}) form the general solution of the two-body problem.

Equation (\ref{equation_eight}) generates a $(z +1)$ square matrix, where $z$ is the coordination number of the lattice. Hence, we obtain a ($9\times 9$) matrix for BCC lattices. It is useful to apply a symmetrized approach which reduces the matrix size as we will demonstrate here. The symmetrization improves numerical stability and speed, and allows for better classification of the results.

The two-particle wave function has to be symmetric or anti-symmetric under spatial exchange $\Psi(\nvec_{1},\nvec_{2})=\pm \Psi(\nvec_{2},\nvec_{1})$ which translate to singlet ($+$) and triplet ($-$) spin states. To express the symmetrized wave functions, we permute $\kvec_{1} \rightarrow \kvec_{2}$ in (\ref{equation_four}) and then we add/subtract the resulting equation from the unpermuted version of Equation (\ref{equation_four}). Doing this yields:

\begin{align}
& (E-\varepsilon_{\kvec_{1}}-\varepsilon_{\kvec_{2}})(\psi_{\kvec_{1}\kvec_{2}} \pm \psi_{\kvec_{2}\kvec_{1}})\nonumber\\
& =\frac{1}{N}\sum_{\avec\qvec}\hat{V}_{\avec}\;\Big\{e^{i(\qvec-\kvec_{1})\avec} \pm e^{i(\qvec-\kvec_{2})\avec}\Big\}\,\psi_{\qvec,\kvec_{1}+\kvec_{2}-\qvec}
\label{equation_eleven}
\end{align}

\noindent We can rewrite the (anti-)symmetrized pair wave functions on the left-hand-side of Equation (\ref{equation_eleven}) as
\begin{equation}\label{equation_twelve}
\phi_{\kvec_{1}\kvec_{2}}^{\pm}=\psi_{\kvec_{1}\kvec_{2}} \pm \psi_{\kvec_{2}\kvec_{1}} ,
\end{equation}
where $\phi^+_{\kvec_{1}\kvec_{2}}$ and $\phi^-_{\kvec_{1}\kvec_{2}}$ respectively are the singlet and the triplet wave functions. So, Equation (\ref{equation_eleven}) becomes 

\begin{align}\label{equation_thirteen}
& (E-\varepsilon_{\kvec_{1}}-\varepsilon_{\kvec_{2}})\phi_{\kvec_{1}\kvec_{2}}^{\pm}\nonumber\\
&=\frac{1}{N} \sum_{\qvec\avec}\hat{V}_{\avec}\;\Big\{e^{i(\qvec -\kvec_{1})\,\avec} \pm e^{i(\qvec -\kvec_{2})\,\avec} \Big\}\;\psi_{\qvec,\kvec_{1}+\kvec_{2}-\qvec} 
\end{align} 
 The summation over the lattice vector $\avec$, in Equation (\ref{equation_thirteen}) can be split into two sets ($\{\avec_+\}$ for singlets, and $\{\avec_-\}$ for triplets) which thus allows us to write the right-hand side of the equation in terms of $\phi^\pm$ instead of $\psi$. To do this, we define $\{\avec_+\}$ and $\{\avec_-\}$ to be a set of near-neighbor lattice vectors, and also include the zero vector in the case of singlets:
\begin{align}
    \begin{split}
    \{\avec_{+}\} & = \{ (0,0,0), (\frac{b}{2},\frac{b}{2},\frac{b}{2}), (-\frac{b}{2},\frac{b}{2},\frac{b}{2}), (\frac{b}{2},-\frac{b}{2},\frac{b}{2}), \\ & \;\;\;\;\;\;\;(\frac{b}{2},\frac{b}{2},-\frac{b}{2})\}
    \end{split}
    \\
\{\avec_{-}\} 
& = \{ (\frac{b}{2},\frac{b}{2},\frac{b}{2}), (-\frac{b}{2},\frac{b}{2},\frac{b}{2}), (\frac{b}{2},-\frac{b}{2},\frac{b}{2}), (\frac{b}{2},\frac{b}{2},-\frac{b}{2})\} 
\label{schrodinger_equation_trip}
\end{align}
where $b$ is the lattice constant. It is important that neither set contains pairs of members that are related by inversion, but otherwise there is some freedom in the choice of selecting the new vectors. Then,
\begin{align}\label{equation_sing_trip_split}
& (E-\varepsilon_{\kvec_{1}}-\varepsilon_{\kvec_{2}})\phi_{\kvec_{1}\kvec_{2}}^{\pm}\nonumber\\
&=\frac{1}{N} \sideset{}{'}\sum_{\qvec\avec_{\pm}}\hat{V}_{\avec_{\pm}}\;\Big\{e^{i(\qvec -\kvec_{1})\,\avec_{\pm}} \pm e^{i(\qvec -\kvec_{2})\,\avec_{\pm}} \Big\}\;\phi^{\pm}_{\qvec,\kvec_{1}+\kvec_{2}-\qvec} 
\end{align} 
Note that the primed summation in Equation (\ref{equation_sing_trip_split}) above means a factor of $\frac{1}{2}$ should be included for the case $\avec_{+}=\mathbf{0}$. Following similar steps in Equations (\ref{equation_six})$-$(\ref{equation_nine}), we obtain

\begin{equation}\label{equation_seventeen}
\Phi_{\avec_{\pm}}^{\pm}(\Pvec)=- \sum_{\avec_{\pm}^{'}}\hat{V}_{\avec_{\pm}}
L_{\avec_{\pm}\,\avec_{\pm}^{'}}^{\pm}(E,\Pvec)\;\Phi_{\avec_{\pm}^{'}}^{\pm}(\Pvec)
\end{equation}
where,
\begin{equation}\label{equation_eighteen}
L_{\avec_{\pm}\,\avec_{\pm}^{'}}^{\pm}(E,\Pvec)=\frac{1}{N}\sum_{\qvec}\frac{e^{i\qvec(\avec_{\pm}-\avec_{\pm}^{'})}\pm e^{i[\qvec\avec_{\pm}-(\Pvec-\qvec)\avec_{\pm}^{'}]}}{-E+\varepsilon_{\qvec}+\varepsilon_{\Pvec-\qvec}}
\end{equation}

\noindent Equations (\ref{equation_thirteen}) to (\ref{equation_eighteen}) are used to obtain the (anti)-symmetrized solutions. Results were validated using second-order perturbation theory and a Quantum Monte Carlo (QMC) code.

\section{Results} \label{sec:pair_properties}

In this section, we study the properties of pairs in the UV model on the BCC lattice including the total energy, pairing diagram, dispersion, pair mass, radius and finally we estimate the BEC transition temperatures. We focus our attention mainly on the $s$-states. However, we will briefly discuss other pairing symmetries ($p$-, $d$- and $f$- states).

\subsection{Total Energy}\label{sec:total energy}

\begin{figure}[h!]
    \centering
	\includegraphics[scale=0.5]{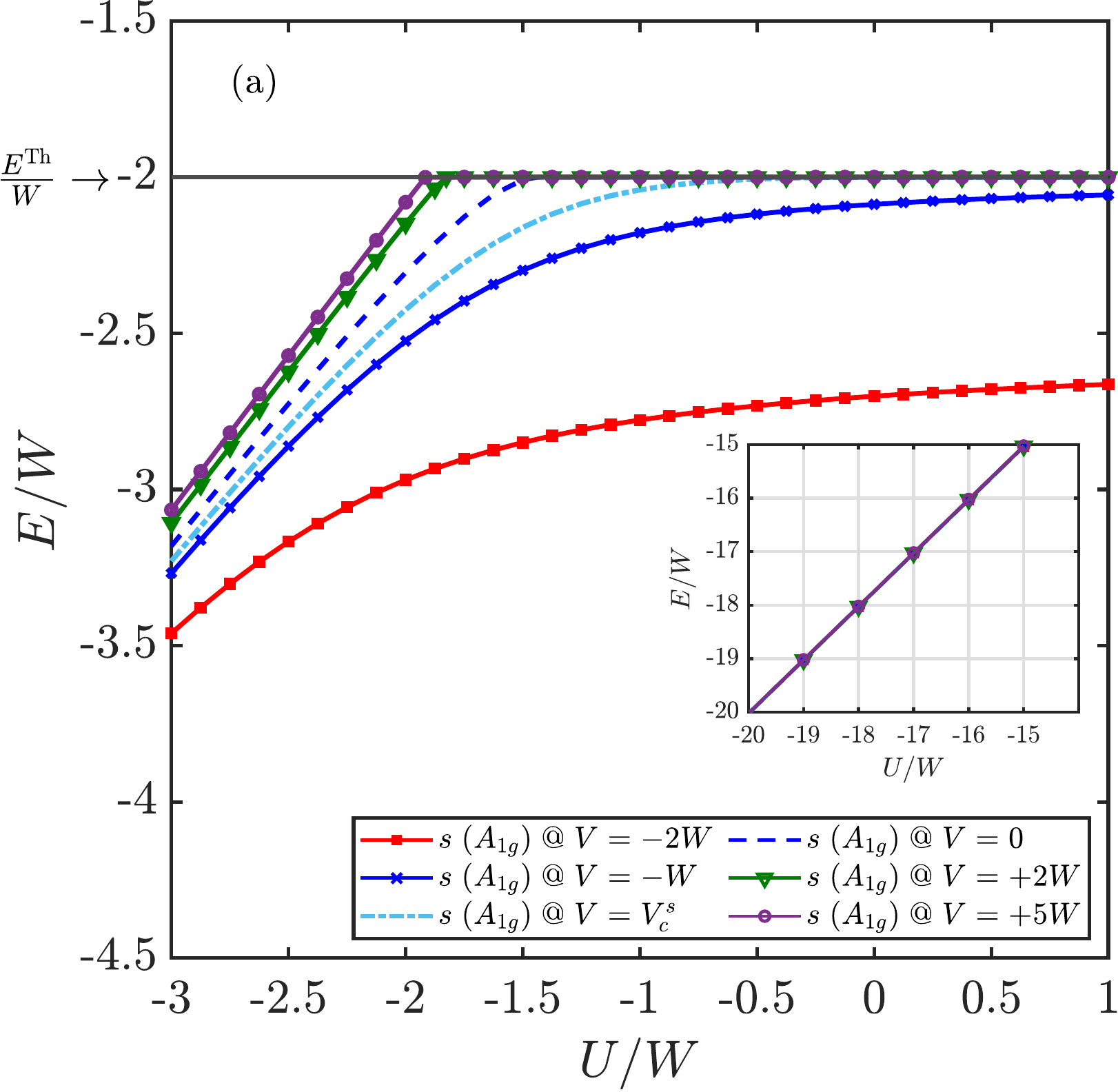}
	\vspace{0.8em}
	\includegraphics[scale=0.5]{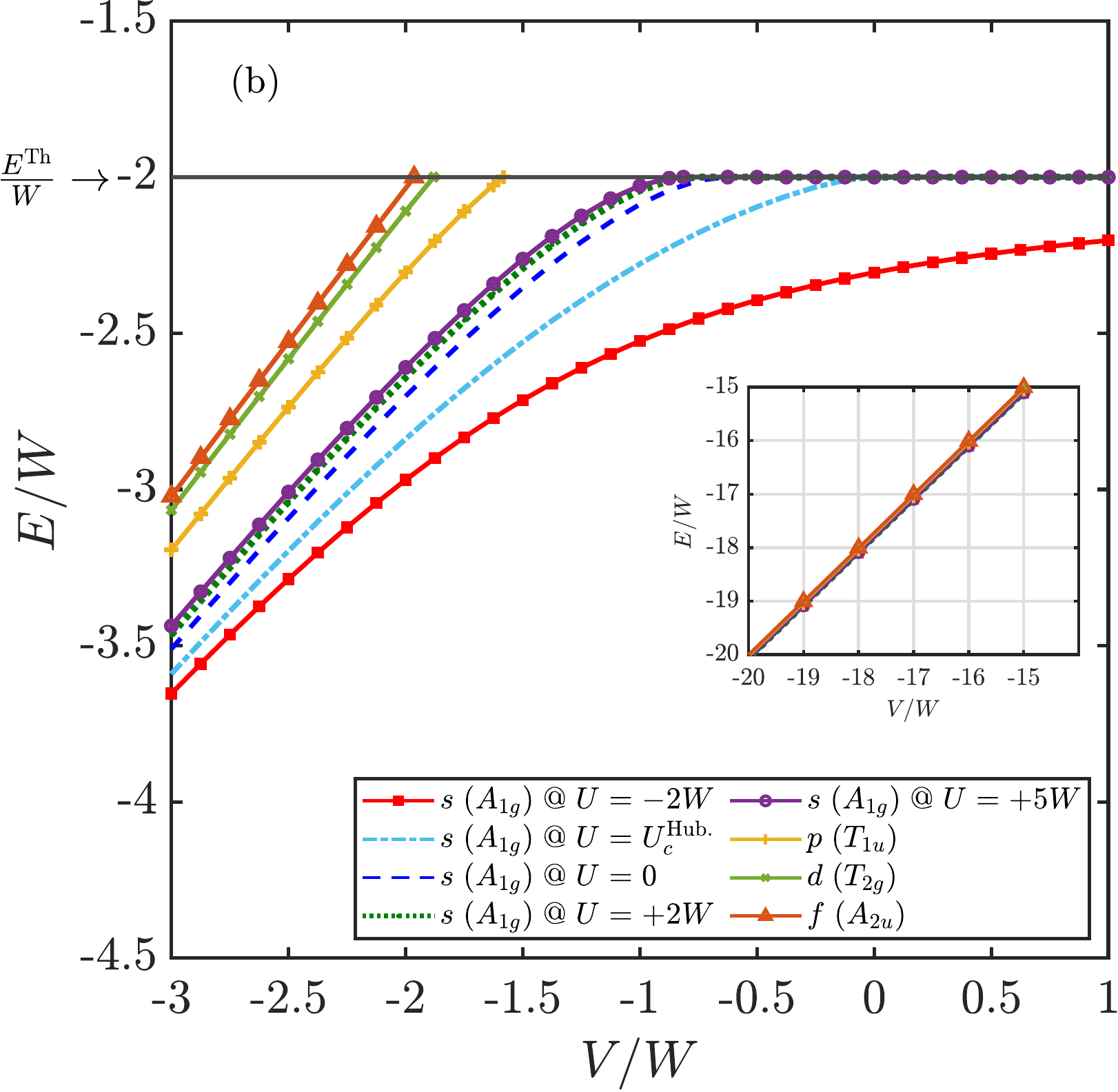}
	\caption{The total energy of pairs. The critical Hubbard attraction for $V=0$ is $U^{\rm Hub.}_{c}(0)= -1.4355W$ and for $U=0$ $V^{s}_{c}(0)= -0.6358W$. The $p$-, $d$- and $f$- states are independent of $U$ i.e. change in $U$ only affects the total energy of the $s$-state. For large onsite, $|U|\gg|V|,t$ (intersite, $|V|\gg|U|,t$) attraction, $E\rightarrow-|U|$ ($E\rightarrow-|V|$) for all the states (inset). The corresponding symmetry of each state is also indicated.}
	\label{fig:total_energy}
\end{figure}

The transition from an unbound (two free particles) to a bound state occurs at a critical value of $U$ and $V$ which can be identified using the pair energy. Figure \ref{fig:total_energy} shows plots of the total energy for different pair symmetries. The flat region of the curve corresponds to the total energy of two unbound particles (the threshold energy $E^{\rm Th}=-2W$, where $W=8t$ is the half-bandwidth). The energy drops below $-2W$ as the attraction gets stronger indicating that a bound pair has been formed. Pairs are highly stable (well bound) at large attractive coupling. The $p$- and $d$-states are both three-fold degenerate, and $s$- and $f$-states have degeneracy 1.

Fig. \ref{fig:total_energy}a and \ref{fig:total_energy}b respectively, show a shift in the critical $U$ ($V$) required to form stable $s$-symmetric pairs when modifying the intersite (onsite) repulsion. A stable $s$-symmetric pair is guaranteed to form if $U\leq-2W$ or $V\leq-0.8858W$ (more details in Sec. \ref{sec: binding diagram} and in the Appendix). For infinite attractive $V$, the particles form deep, localized pairs and the energies associated with all the intersite pairing symmetries converge, i.e. $E \rightarrow -|V|$ (inset plots in Fig. \ref{fig:total_energy}). For non-$s$ pairing symmetries, the critical interaction is independent of $U$. For the onsite $s$-pair, $E\rightarrow -|U|$ as $U\rightarrow -\infty$. In these large interaction limits, binding energies and thus binding temperatures can be large.

\begin{figure}[t!]
	\centering
	\includegraphics[width=92mm]{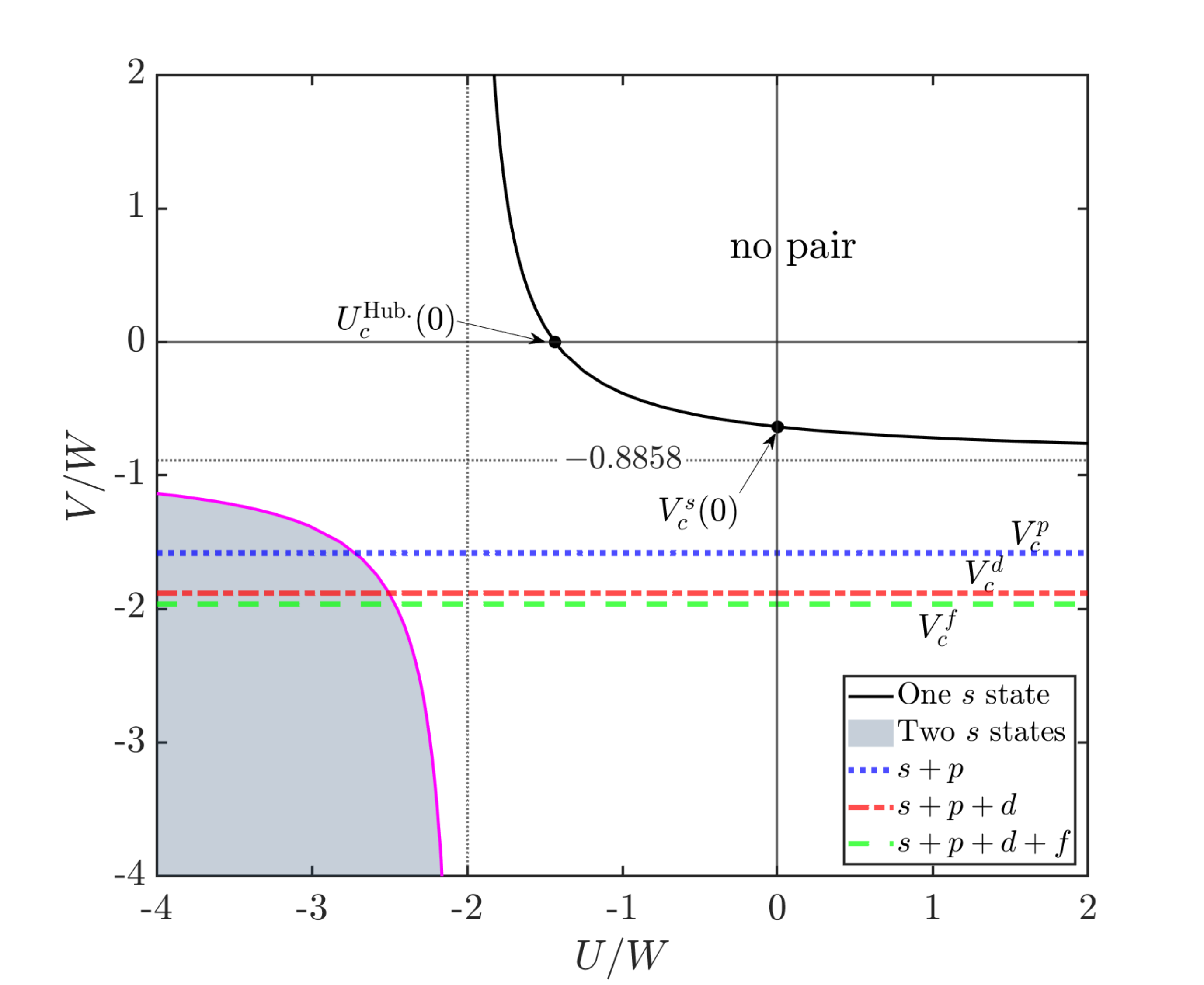}
	\caption{[Color online] Binding diagram for pair formation when $\Pvec=0$ (temperature $T=0$) on the BCC lattice. The top (curved) solid line shows the formation of one bound $s$-state pair (singlet), the shaded region enclosed by the (magenta) solid line indicates region of formation of two $s$-states, the (blue) dotted line shows the onset of triply degenerate $p$-states (three $p$-wave triplets), the (red) dash-dotted line shows the binding of the three-fold degenerate $d$-states (three $d$-wave singlets) and the (green) dashed line indicates the formation of a pair with $f$-symmetry (one $f$-wave triplet). The $p$-, $d$- and $f$- states begin to appear at critical intersite attractions $-1.5828W$, $-1.8803W$ and $-1.9639W$ respectively. The smaller dotted vertical and horizontal lines are the asymptotes ($V_{\rm asym.}^s\approx-0.8858$W and $U_{\rm asym.}^{\rm Hub.}=-2W$) and for potentials equal to or more attractive than these values, the creation of an $s$-state bound pair on the BCC optical lattice is guaranteed.}
	\label{fig:bcc-binding-diagram}
\end{figure}

\subsection{Binding Diagram}\label{sec: binding diagram}

By identifying the point at which the total energy drops below $-2W$, the binding diagram at $\Pvec=0$ can be constructed (Fig. \ref{fig:bcc-binding-diagram}). A pair of free, unbound particles with zero total momentum has energy $E=-2W$. So, the threshold energy is $E^{\rm Th}=-2W$. For any pairing symmetry with a node at the origin, binding is independent of $U$.

Within the $U$-$V$ parameter space, pairing is found at large, attractive $U/W$ and/or $V/W$. The kinetic energy of the particles on the BCC optical lattice is large (relative to 1D, 2D and simple cubic lattices) due to the larger coordination number and, as a result, $U$ or $V$ must be large and attractive in order to form a bound state. The critical binding values $U_c$ and $V_c$ are derived in the Appendix. The critical attraction at $\Pvec=0$ can be determined via 
\begin{equation}
    V_{c}^{s}(U)\leq \frac{UL_{0}-1}{UL_{0}\mathcal{C}-\mathcal{C}-8UL_{1}^2}\;\;,
\end{equation}
where $L_0$=$-K_{0}^2/(4\pi^2t)$, $L_1$=$ L_0 + 1/(16t)$, $\mathcal{C}$=$8L_0 + 1/(2t)$ and $K_0=K(1/\sqrt{2})$ is the complete elliptic integral of the first kind. For a negative-$U$ Hubbard model with no intersite interaction, the critical binding is found to be $U_{c}^{s}$($V$=0)$\approx$$-1.4355W$. Similarly, $V_{c}^{s}$($U$=0)$\approx$$-0.6358W$ is required to bind particles when the onsite interaction is absent. As noted in Section \ref{sec:total energy}, an intersite strength $V^s_{c}(+\infty)$$\approx$$-0.8858W$ is sufficient to maintain a bound state even if the Hubbard repulsion is infinite while $U^{s}_{c}$($V\rightarrow +\infty$)=$-2W$. Figure \ref{fig:bcc-binding-diagram} also shows the binding thresholds of the $p$-, $d$- and $f$-states respectively occurring at quite large intersite attractions, i.e. $V_{c}^p$=$-1.5828W$, $V_{c}^{d}$=$ -1.8804W$, $V_{c}^{f}$=$-1.9639W$.

\begin{figure}[h!]
    \centering
    \includegraphics[width=85mm]{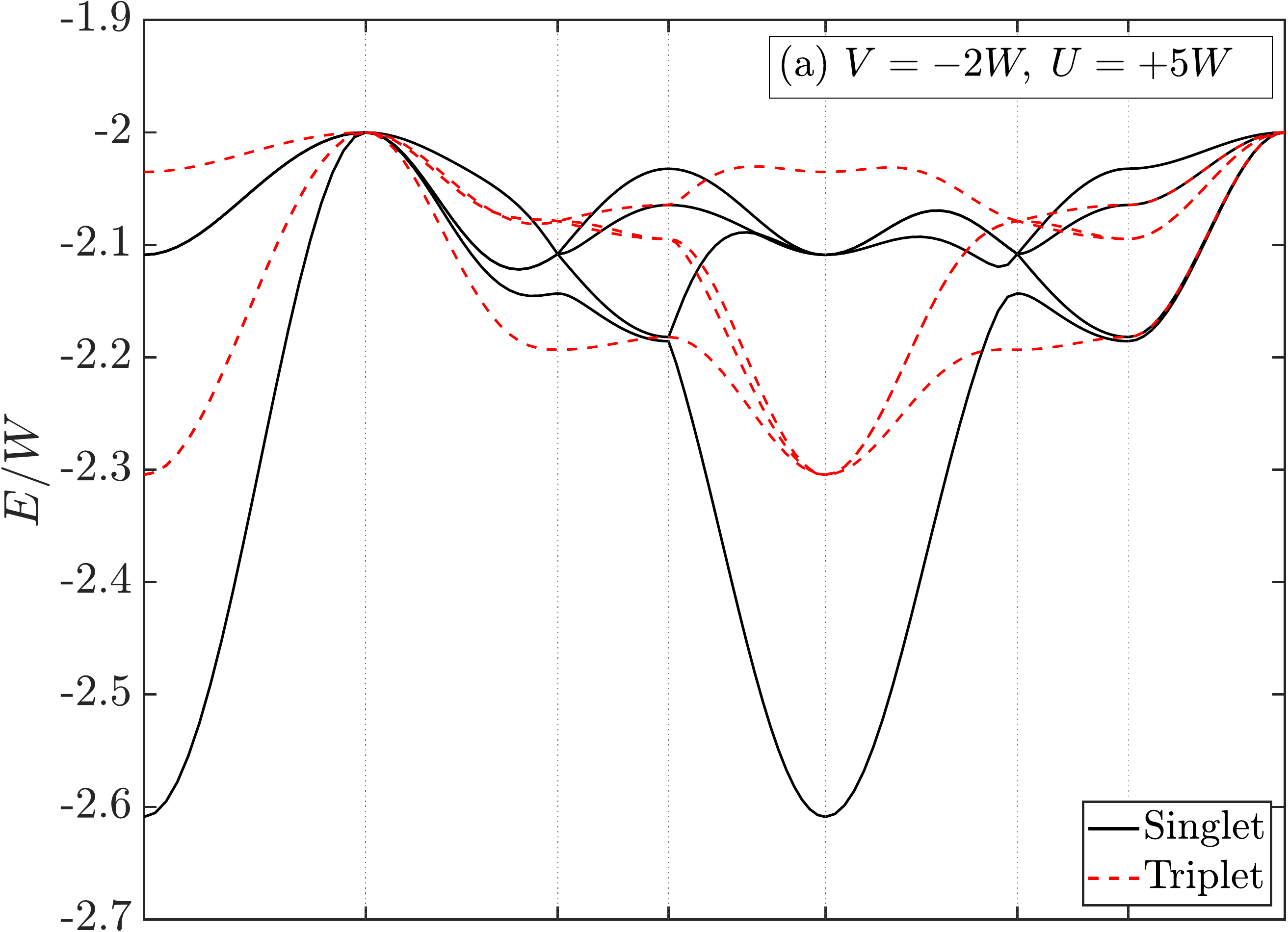}
    \includegraphics[width=85.mm]{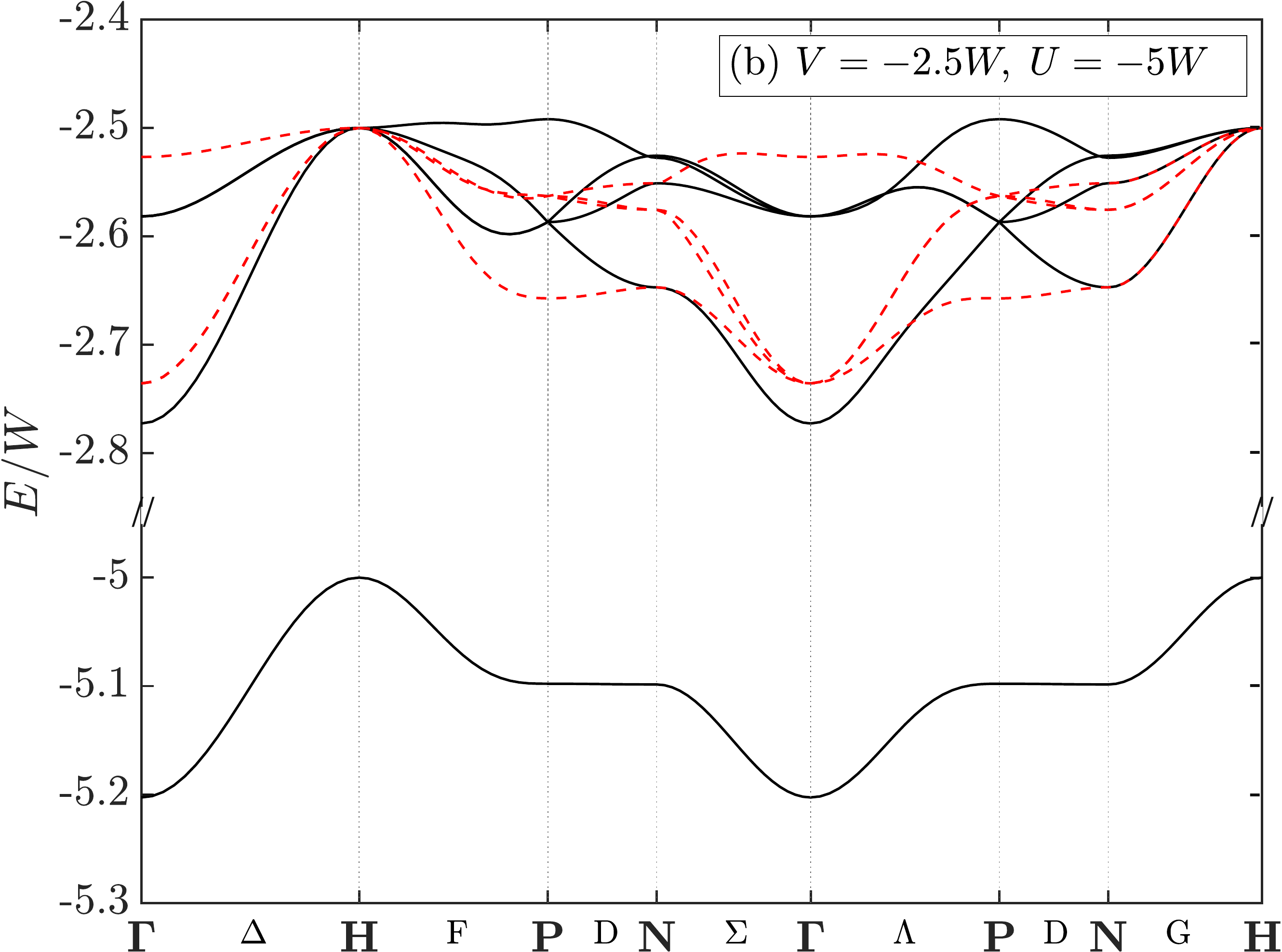}
    \caption{Examples of dispersions with (a) repulsive $U$ and attractive $V$, (b) $U$ and $V$ are both attractive. Solid (dotted) lines are the singlet (triplet) states. In (b) there is a low-lying onsite state. Note the break in the y axis in (b).}
	\label{fig:bcc_dispersion_plot}
\end{figure}

\begin{figure}[h!]
    \centering
    \includegraphics[width=85mm]{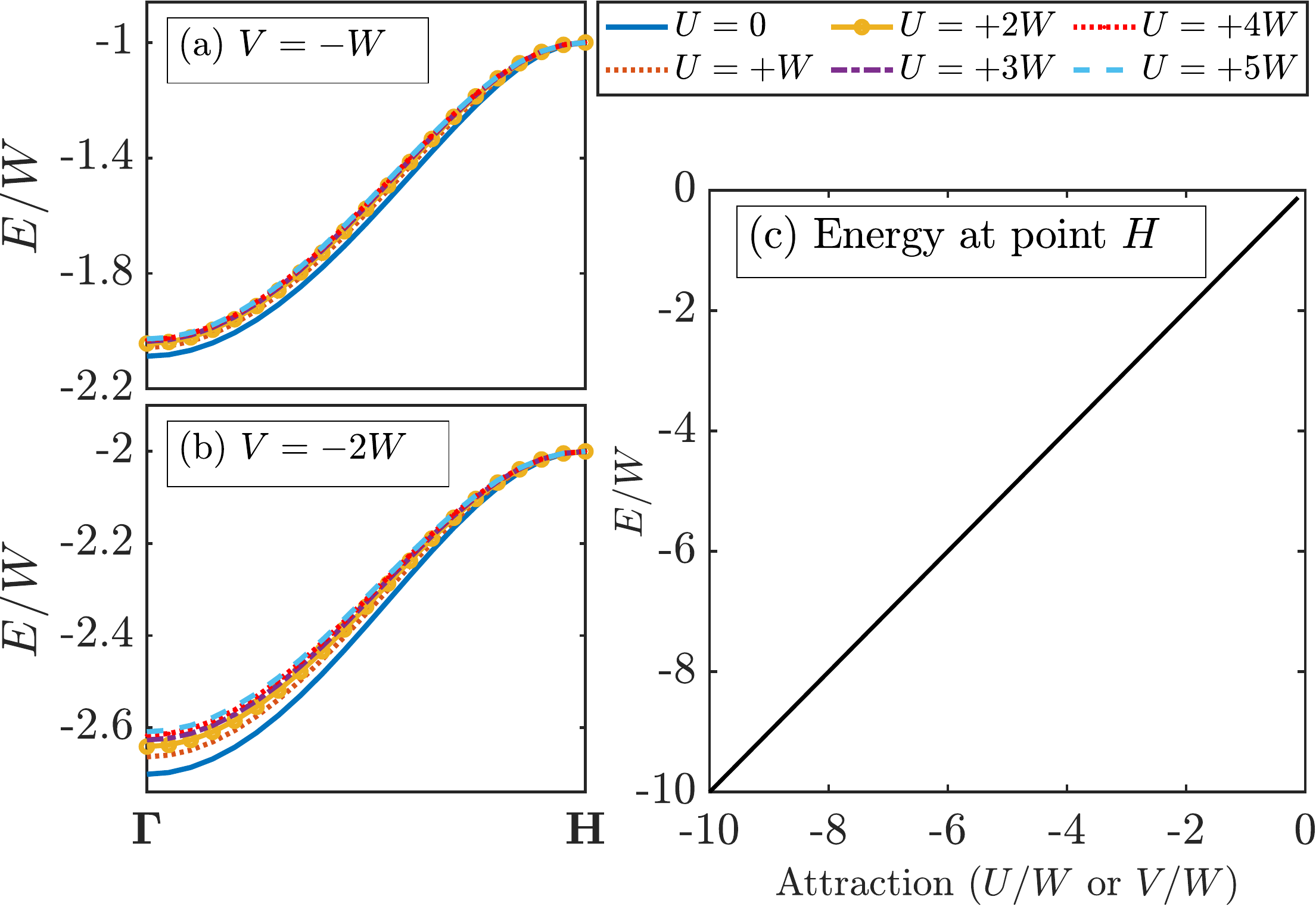}
    \caption{[Color online] Panels (a) and (b): Dispersion of the extended $s$-state only on the $\Gamma$-$H$ line showing that it is independent of repulsive $U$ at the $H$ point. Panel (c): energy of all bound states at point $H$ (N.B. extended-$s$, $d$, $p$ and $f$ states are only bound if $V$ is attractive, and the onsite $s$ state is only bound if $U$ is attractive). This shows that a pair, however the symmetry, can be formed at very weak attraction, in contrast to the critical attractions $U_c, V_c$ (refer to Section \ref{sec: binding diagram}, Fig. \ref{fig:bcc-binding-diagram}), required at the $\Gamma$ point.}
    \label{fig:h-point_different_U}
\end{figure}

\subsection{Dispersion}
Estimation of the pair condensation temperature requires the pair effective mass at the $\Gamma$ point, for which it is necessary to calculate the dispersion (pair energy at non-zero momentum). Examples of the dispersion for various $U$ and $V$ are presented in Fig. \ref{fig:bcc_dispersion_plot}. The dispersion for two free particles is not shown but lies in the range $-2\leq E/W \leq 2$. The singlet $s$-symmetric pair has the lowest energy around the Brillouin zone (BZ) center (i.e. $\Gamma$ point) but this is not the case at other high symmetry points. The band structure gets narrower as the intersite attraction $V$ increases, consistent with an increase in the effective mass. There is a high level of degeneracy on the $\Gamma$-$H$ line; all states with the same symmetry class are degenerate and the simplification at the $\Gamma$ point (see Appendix) applies too. The momentum contribution only renormalizes the hopping parameter $t$.

Away from the $\Gamma$-$H$ line, states with different symmetries mix, although the singlet and triplet states can be unambiguously classified because the (anti-)symmetrized solution separates singlets from triplets. From the dispersion calculations, we note that away from the $\Gamma$ point, there is a possibility of forming bound pairs with vanishingly small attraction. This provides further insights into pair stability at different attractive $U$ and $V$ values. Across the BZ, there are singlet-triplet crossings especially along $N$-$H$. At the $H$ point where the total pair momentum is maximal, there is some special behavior. Firstly, the bound onsite s-state has an energy equal to $U$ and all other states (which are bound by the intersite potential) have energy $V$ (as long as the relevant potential is non-zero and attractive). Secondly, for repulsive $U$, the extended-$s$ state becomes completely independent of the Hubbard $U$ repulsion  (Fig. \ref{fig:h-point_different_U}, panels (a) and (b)). Thirdly, a vanishingly small attraction ($U\rightarrow 0^{(-)}$ and $V\rightarrow 0^{(-)}$) is sufficient to bind pairs (Fig. \ref{fig:h-point_different_U}c).

\begin{figure}[h!]
	\centering
	\includegraphics[width=85mm]{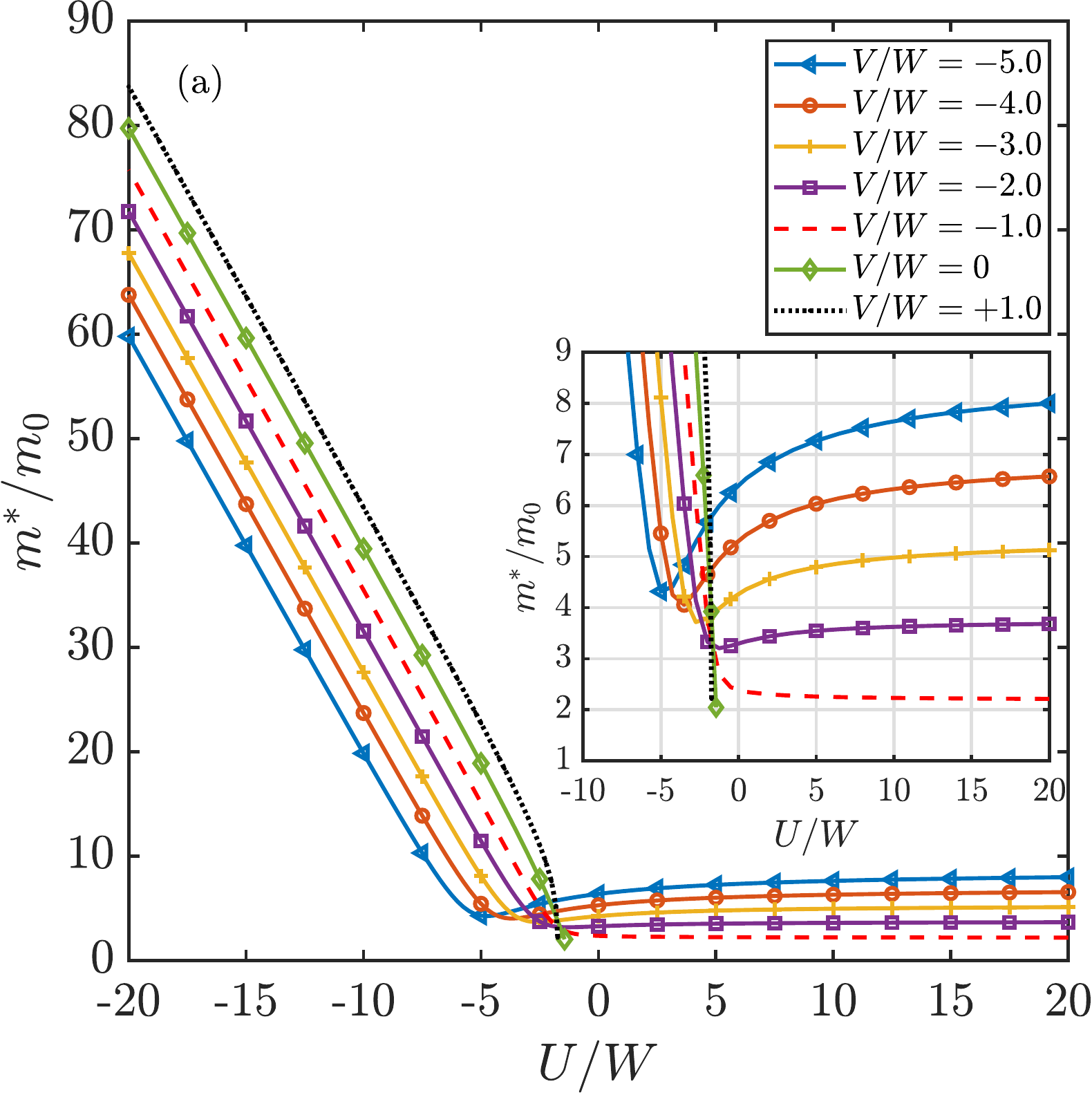}\vspace{.08em}
	\includegraphics[width=85mm]{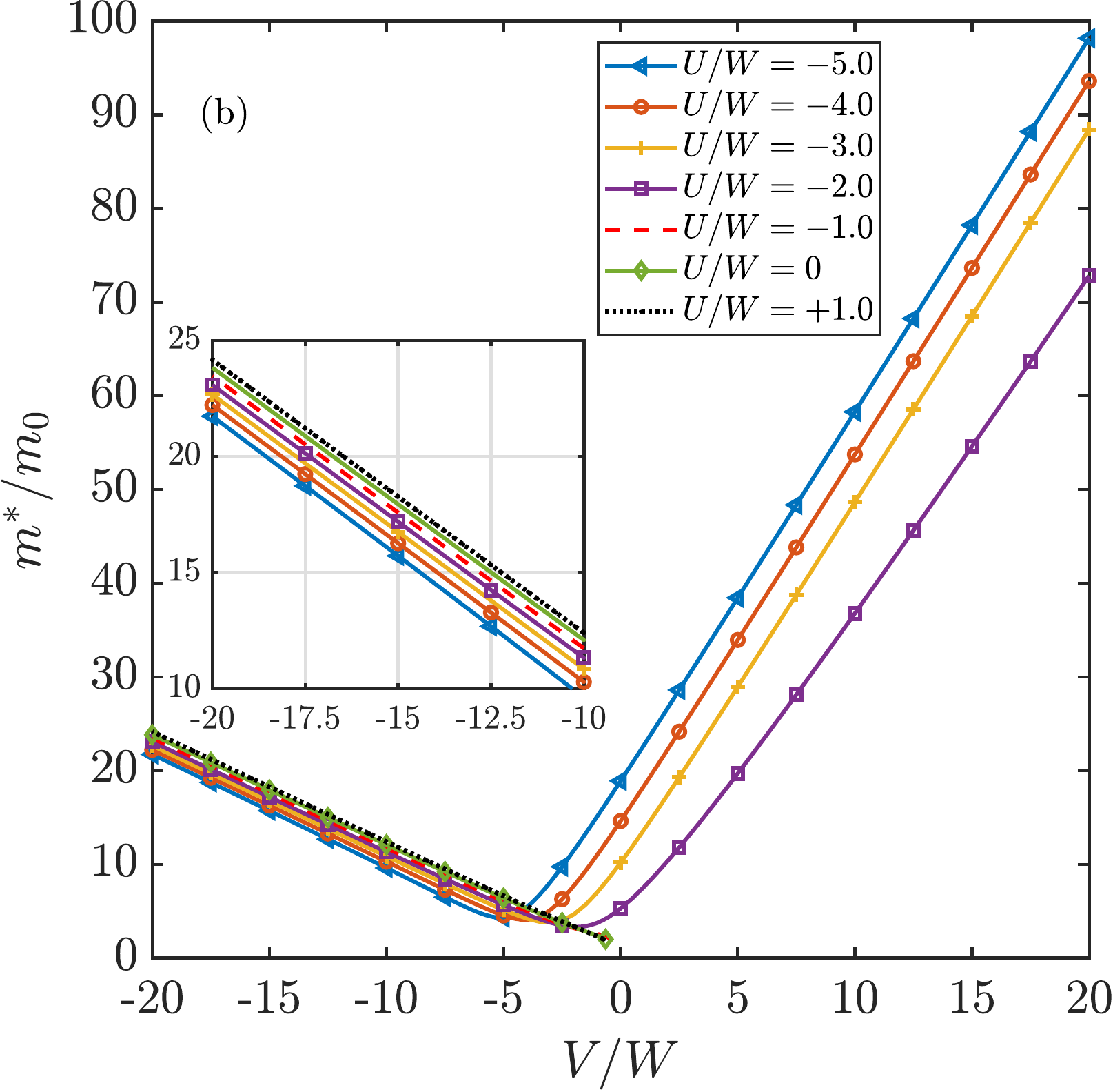}
	\caption{[Color online] The pair mass expressed in unit of a free particle mass. $m_{0}=\hbar^2/(2b^2t)$ is the bare mass of one free particle on the BCC optical lattice. The value of $m_{0}$ will be discussed in Sec. \ref{sec:transitiontemperature}.}
	\label{pairmassbcc}
\end{figure}

\begin{figure}
    \centering
    \includegraphics[width=85mm]{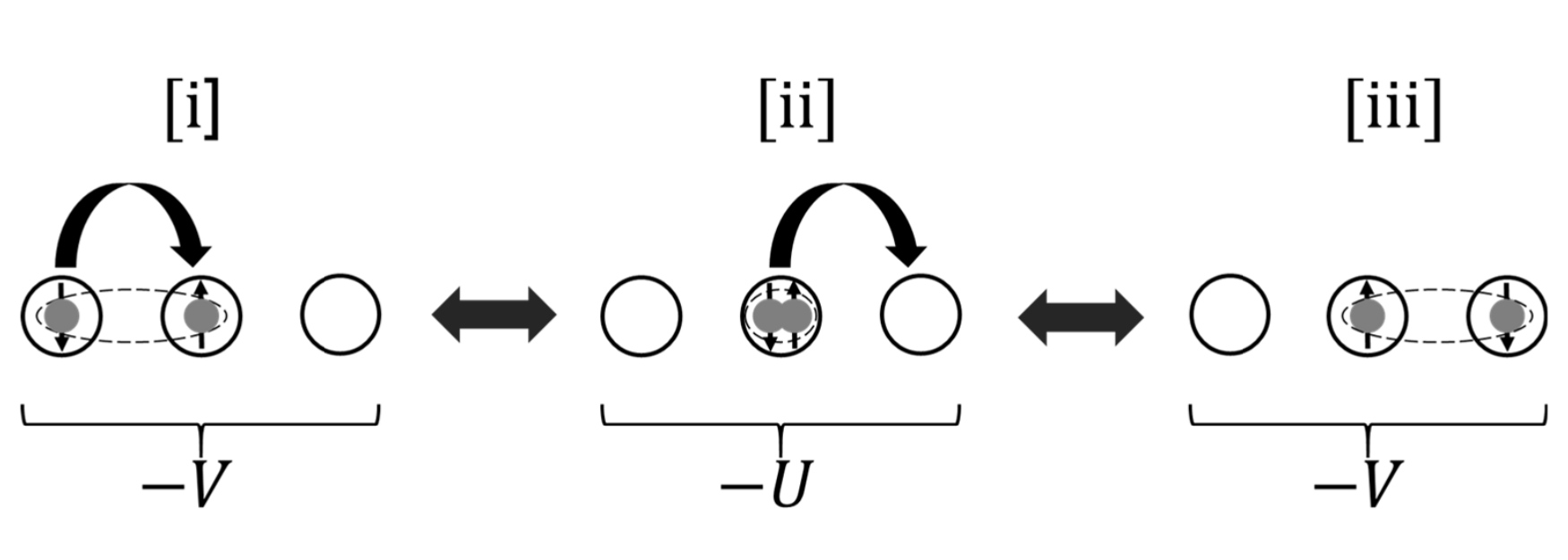}
    \caption{One-dimensional schematic of superlight behavior of a singlet when $U$ and $V$ are comparable and sufficiently attractive. The big circle, gray ball (with vertical arrow) and the dashed-line oval represents lattice site, atom and bonding respectively. The two-way arrow indicates that the total energy of the pair, $E$, is the same, thus switching between configurations comes with no energy penalty.}
    \label{fig:superlight_pair_hopping}
\end{figure}

\begin{figure}
	\centering
		\includegraphics[width=85mm]{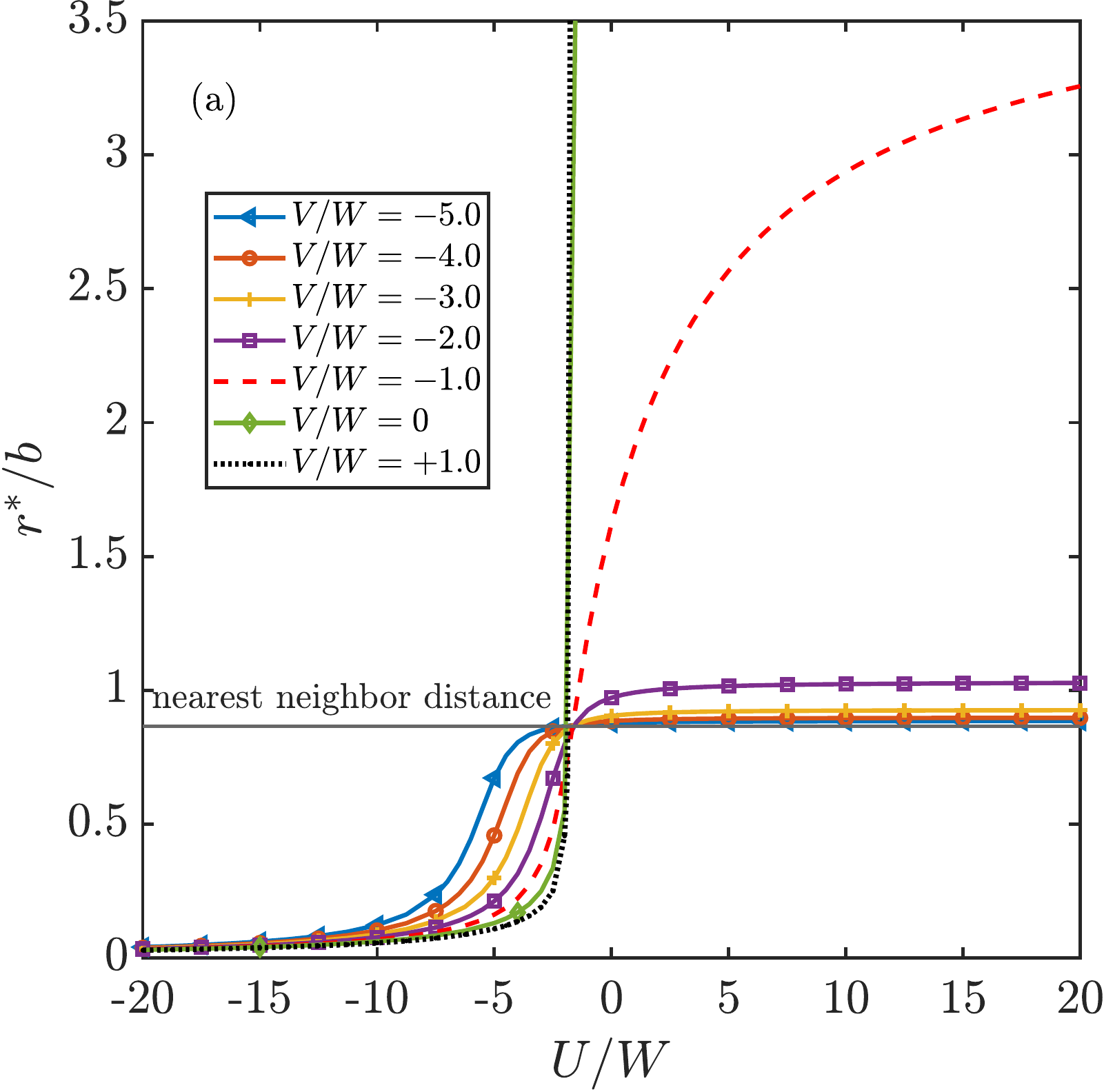}    \includegraphics[width=85mm]{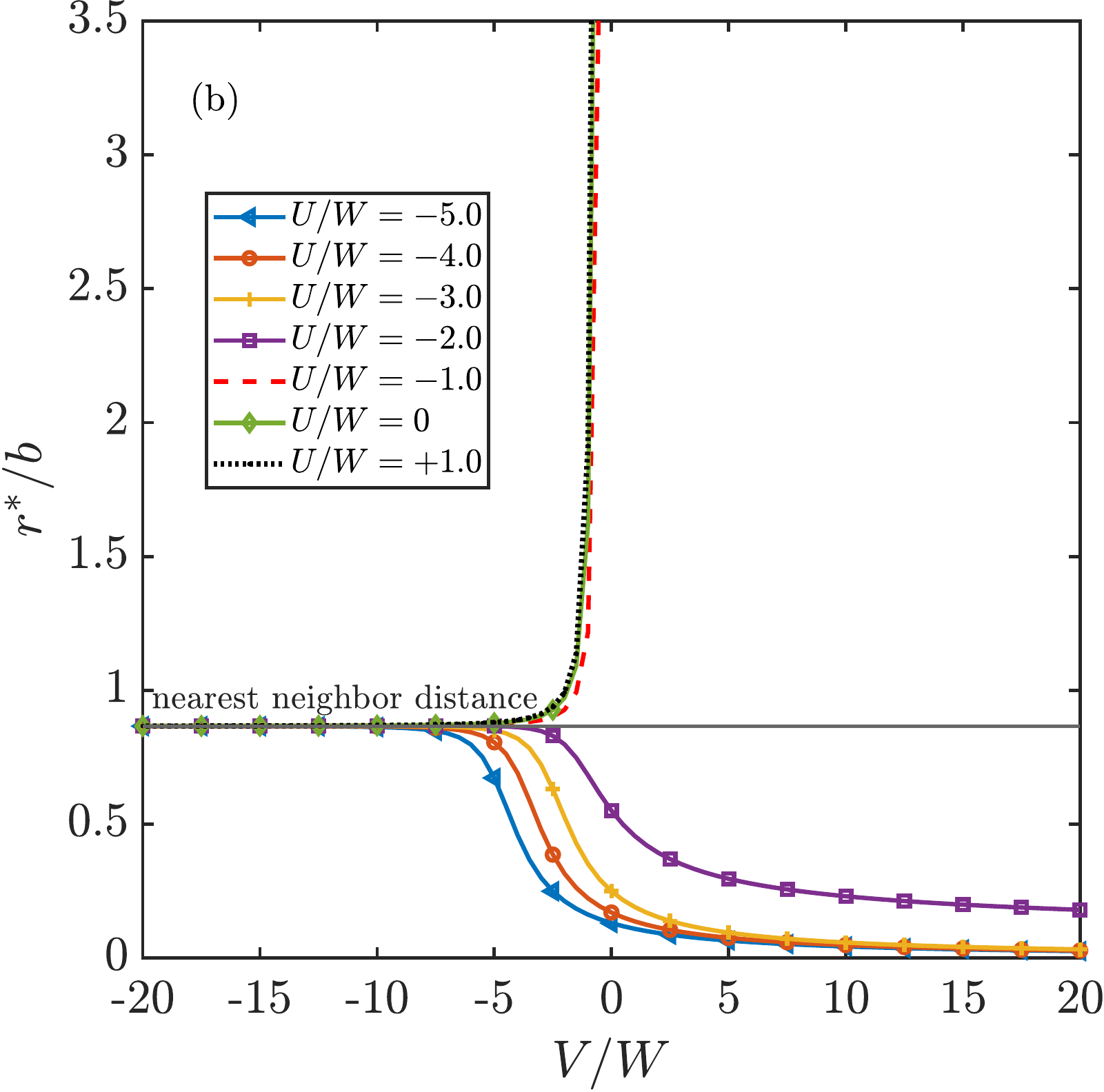}
	\caption{[Color online] Effective radius for various $U$ and $V$. The thin solid horizontal lines represent the nearest neighbor distance $a=\sqrt{3}b/2$ on a BCC lattice. Note that the radius diverges at low attraction. A shoulder forms in the curves when $U\sim V$.}
	\label{fig:pair size bcc}
\end{figure}

\subsection{Pair Mass}

Pair mass can be calculated from the second derivative of the dispersion within the standard effective mass approximation, as,
\begin{equation}
    [m^{*}_i]^{-1}=\frac{1}{\hbar^2}\frac{\partial^2 E}{\partial P_{i}^2}. 
\end{equation}
The resulting effective masses can be seen in Fig. \ref{pairmassbcc}.

Weakly bound pairs are light (about $2m_0$ when pair energy approaches the threshold energy $E^{\rm Th}$) and get heavier as the binding attraction is increased. (Note that $m_0=\hbar^2/(2b^2t)\,$ is the bare effective mass of a free particle.)

Pairs with strong intersite attraction are much lighter in comparison to their counterpart pairs with strong onsite attraction (for example, compare the masses at $U=-20W$ in Fig. \ref{pairmassbcc}(a) and at $V=-20W$ in Fig. \ref{pairmassbcc}(b)). This occurs since large attractive $U$ confines the pair to the same site thereby suppressing the probability of particles hopping to other sites (the suppression is even greater when $V$ is repulsive).

When $U$ and $V$ are both attractive and of similar magnitude, we find superlight pairs (minima in Figs. \ref{pairmassbcc}a and \ref{pairmassbcc}b). These correspond to the situation where particles can move without passing through a high energy intermediate state. Figure \ref{fig:superlight_pair_hopping} demonstrates this superlight process on the optical lattice where the pair travels through the lattice in a crawler motion. The process of movement can be summarised in three main steps: (i) the particles interact via attractive $V$; (ii) one particle hops to interact with the other via an attractive $U$; (iii) either of the particles can hop so interaction is intersite again. While it moves around, the pair's total energy remains unchanged in all the configurations.

\subsection{Pair Radius}

The effective radius has been calculated using the relation
\begin{equation}\label{pair_radius_equation}
    \langle r^*\rangle=\sqrt{\frac{\sum_{\nvec}\nvec^2\Psi^*(\nvec_{1},\nvec_{2})\Psi(\nvec_{1},\nvec_{2}) }{\sum_{\nvec}\Psi^*(\nvec_{1},\nvec_{2})\Psi(\nvec_{1},\nvec_{2})}} \;\;\;,
\end{equation}
where $\nvec=\nvec_{1}-\nvec_{2}$ is the spatial separation between the particles and $\Psi(\nvec_{1},\nvec_{2})$ is the pair wave function.

Figure \ref{fig:pair size bcc} shows the pair radius. Near the threshold energy ($E\rightarrow E^{\rm Th}$), the particles form a large pair: a consequence of the delocalization of the pair wave function. At intermediate $U$ and $V$ (both attractive), the pair's size is on the order of the near-neighbor distance $a$ (the horizontal line in Fig. \ref{fig:pair size bcc}). In a pairing scenario where $V$ is fixed and $U$ is tuned to be highly attractive, the two bound particles are localized and held on the same site. The pair is also local for large intersite attraction but the size levels off to the nearest neighbor distance at large attractive $V$. By local, we mean bound pairs that are not larger than the lattice constant.

\begin{figure*}%
	\centering
	(i) Fixing $V$ and tuning $U$\\
	\includegraphics[width=152mm]{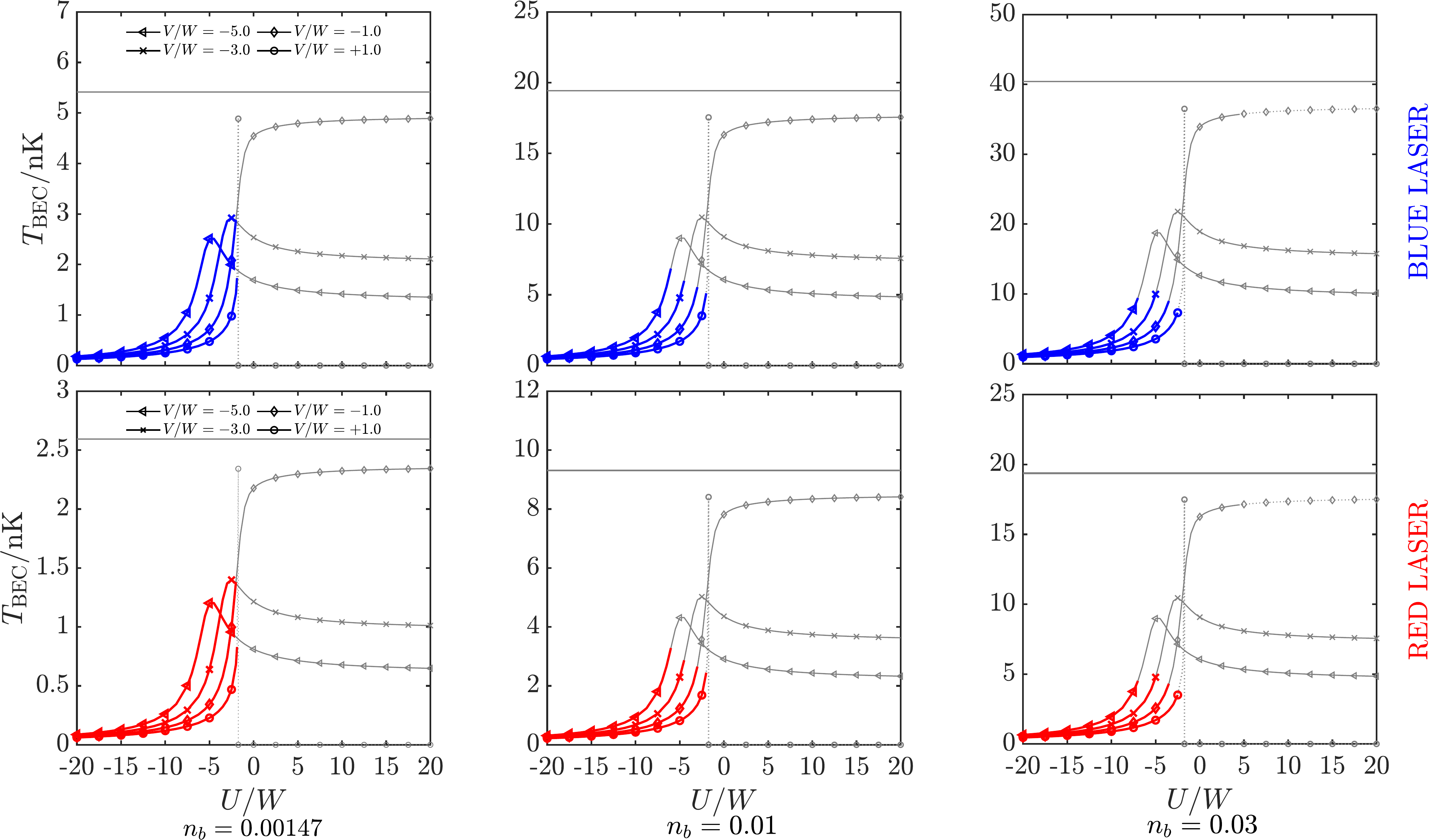}
	
       (ii) Fixing $U$ and tuning $V$\\
    \includegraphics[width=152mm]{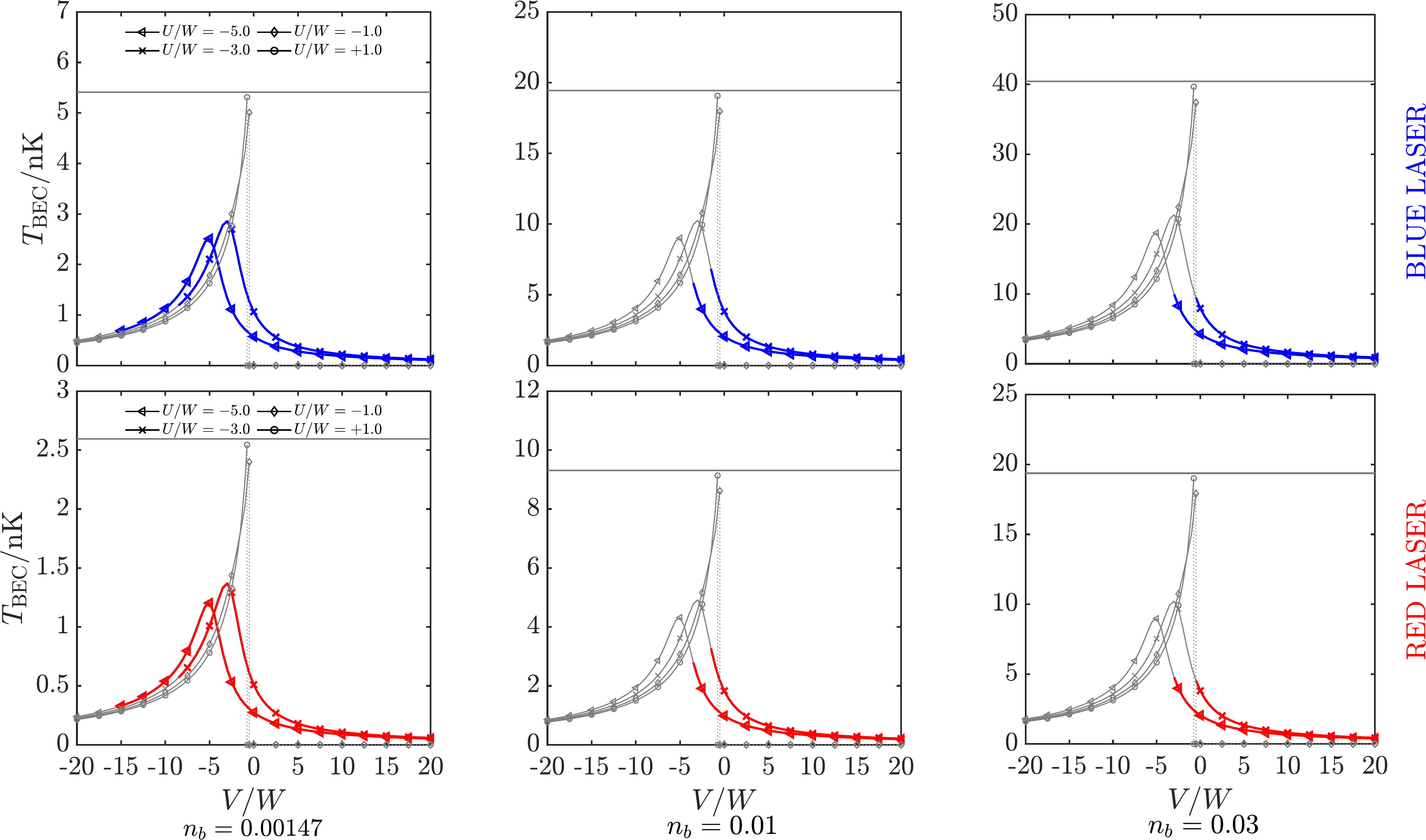}  
	\caption{[Color online] BEC transition temperatures, $T_{\rm BEC}$, for $^{6}\rm Li$ atoms. Rows are labeled according to the laser color, i.e. either a red or blue beam laser. The number of pairs per site $n_b$ increases from left to right and the (colored) dark regions of the plots indicate cases where the value of $n_{b}$ is compatible with the conditions on pair overlap. The horizontal lines in each panel are the corresponding $T_{\rm BEC}$ value for $m^* = 2m_{0}$. The dotted regions imply that $T_\Delta < T_{\rm BEC}$ i.e. the pair is formed below the condensation temperature.  $T_{\rm BEC}$ peaks when $U=V$ and the pairs are superlight.}
\label{fig:transition_plot_blue_and_red_laser}
\end{figure*}

\subsection{Transition Temperature}
\label{sec:transitiontemperature}

Bose-Einstein condensation of the pairs may take place on the optical lattice for well-separated, weakly-interacting, local pairs at low temperature (i.e. a dilute system). For a system of pairs within a lattice, the BEC transition temperature can be calculated from the Bose integral:
\begin{equation}
    \frac{n_{b}}{\Omega_{\rm site}}=\int \frac{d^3\Pvec}{(2\pi)^3}\frac{1}{\exp[(E_{\Pvec}-E_{0})/k_{B}T_{\rm BEC}]-1}
\end{equation}
where the pair dispersion is used in this expression. At low energies, the pair dispersion is parabolic, as can be seen in Fig. \ref{fig:bcc_dispersion_plot}.

The Bose--Einstein distribution decreases rapidly when $E_{\Pvec}-E_{0}>T_{\rm BEC}$. This means that only $\Pvec$ points where $E_{\Pvec}\lesssim k_{B}T_{\rm BEC}$ contribute to the integral. 
For pair dispersions that are parabolic (i.e. consistent with an effective mass approximation) up to energy scales comparable with $k_{B}T_{\rm BEC}$, the transition temperature is:
\begin{equation}
	T_{\rm BEC} \approx \frac{3.31\hbar^{2}}{m_{b}^{*}k_{B}} \left( \frac{n_{b}}{\Omega_{\rm site}} \right)^{2/3}
		\label{eqn:bectemp}
\end{equation}
where $m_{b}^{*}$ is the pair mass, $n_{b}$ is the number of pairs per lattice site, $\Omega_{\rm site}= b^3/2$ is the volume of the Wigner--Seitz cell for a BCC lattice, and  $k_{B}$ is Boltzmann's constant. Note that $n_{b}$ is the number of pairs per site and should not be confused with the particle density, which is $n_{b}/\Omega_{\rm site}=2n_b/b^3$.

A four laser scheme can be used to construct a BCC optical lattice (see scheme A from Table 1 of Ref. \cite{yuan2003}), for which $b/\lambda = \sqrt{3}/2$. Then $\Omega_{\rm site} = \lambda^{3} 3^{3/2}/16$, so we can re-write (\ref{eqn:bectemp}) in terms of the laser beam wavelength $\lambda$, as
\begin{equation}
	T_{\rm BEC} = 7.01\frac{n_{b}^{2/3}\hbar^{2}}{m_{b}^{*}k_{B}\lambda^{2}}.
	\label{eqn:tbeclambda}
\end{equation}
In the following, we assume either a red laser with wavelength 650 nm or a blue laser with wavelength 450 nm, and fermionic $^{6}\rm Li$. Equations (\ref{eqn:bectemp}) and (\ref{eqn:tbeclambda}) are valid when pairs are dilute and weakly interacting. This means both that (1) $n_{b}$ must be small, and (2) that the pair wave functions should not overlap strongly, otherwise corrections would be needed to these equations. We also require that pairs exist above $T_{\rm BEC}$, for which we require that $T<T_{\Delta}$, where $T_{\Delta}=\Delta/k_{B}$ is a characteristic pairing temperature and $\Delta=2\varepsilon_{0}-E_{0}$ (where $\varepsilon_{0}$ is the energy of a free particle with zero momentum, $E_{0}$ is the energy of a bound pair).

We can estimate the maximum $n_{b}$ for which $T_{\rm BEC}$ is consistent with the effective mass approximation in the following way: We can relate an effective pair hopping from, $t_{\rm eff}$ to the effective mass via $m_{b}^{*}=\hbar^2/2 b^2 t_{\rm eff}$. Substituting in Eq. \ref{eqn:tbeclambda}, we obtain,
\begin{equation}
	T_{\rm BEC} = 10.52 \frac{n_{b}^{2/3}t_{\rm eff}}{k_{B}}.
\end{equation}
Typical pair dispersions have parabolic behavior up to at least $E_{\Pvec}\sim t_{\rm eff}$. We set an upper bound that $k_{B}T_{\rm BEC}\lesssim t_{\rm eff}$ for the effective mass approximation to be reasonable. This leads to the estimate that $n_{B}\lesssim 10.52^{-3/2}=0.0293$ for $T_{\rm BEC}$ to be consistent with the effective mass approximation. 

The hopping in a sinusoidal optical lattice can be estimated following Ref. \cite{bloch2008} to be,
\begin{equation}
    t\approx \frac{h^2}{2\sqrt{\pi}Ma^2}\left(\frac{V_0}{E_{r}}\right)^{3/4}\exp\left[-2\left(\frac{V_0}{E_{r}}\right)^{1/2}\right]
    \label{eqn:hopping}
\end{equation}
where $a$ is the nearest neighbor distance, $V_0$ is the depth of the lattice potential, $E_r$ is the recoil energy and $M$ is the atomic mass of the atoms. According to Ref. \cite{bloch2008} , this is accurate to $\sim$10\% for $V_0/E_{r}\gtrsim 15$, and $\sim$15\% for $V_0/E_{r}\gtrsim 10$ . Note that the recoil energy $E_r$ is defined as
\begin{equation}
    E_{r}=\frac{h^2}{8Ma^2} \hspace{2em}.
    \label{eqn:recoilvsa}
\end{equation}
In this paper, we have used $V_0/E_r=10$. This means that the hopping energy $t$ scales as $\sim10^{-12}$ electron-volt for $^6$Li atoms. We also note that Eqn. (\ref{eqn:hopping}) is vital as it allows one to express the bare mass
\begin{equation}
    m_{0}=\frac{\hbar^2}{2b^2 t}
    \label{eqn:massvshopping}
\end{equation}
in terms of the atomic mass $M$ as 
\begin{equation}
    m_{0}\approx \frac{M a^2}{4 b^2 \pi^{3/2}}\left(\frac{E_r}{V_{0}}\right)^{3/4}\exp\left[2\left(\frac{V_0}{E_{r}}\right)^{1/2}\right]
\end{equation}
For BCC lattice, $a=\sqrt{3}b/2$, which means that $a^2/b^2=3/4$. Hence,
\begin{equation}
   m_{0}\approx \frac{3M}{16\pi^{3/2}}\left(\frac{E_r}{V_{0}}\right)^{3/4}\exp\left[2\left(\frac{V_0}{E_{r}}\right)^{1/2}\right]
\end{equation}
Our condition on overlap is whether pairs of radius $R'=\alpha r^{*}$ and density $n_{B}$ can fit into space, i.e. that $n_{B}8 R^{\prime 3}/3<1$. The constant $\alpha$ should be sufficiently large that overlap of the exponentially decaying tails of the pair wave function are small. We suggest taking $\alpha=5$. At higher densities we expect that pairs will start to interact and overlap and that would lead to many-body corrections that cause the transition temperature to level out and then fall as pairs begin to interact strongly and other phases are encountered (and potentially clusters form \cite{kornilovitchcluster1,kornilovitchcluster2,kornilovitchcluster3}).

We explore how transition temperatures vary for fixed $n_{b}$, showing how there is a peak associated with superlight behavior (Fig. \ref{fig:transition_plot_blue_and_red_laser}). The (colored) dark regions of the plots indicate cases where the value of $n_{b}$ is compatible with the conditions on pair overlap. Also, the dotted regions imply that the binding temperature $T_{\Delta}$ is lower than the estimated condensation temperature $T_{\rm BEC}$, thus the bound state cannot Bose condense. In order to probe the $T_{\rm BEC}$ of non-overlapping pairs with superlight characteristics (the peak in $T_{\rm BEC}$) in an optical lattice setting, the number of pairs has to be low (say $n_{b}\sim 0.0015$). In that case, the superlight behavior occurs at roughly 3 nK.

\section{Discussion and conclusions}\label{sec:discussion}

In this paper, we have investigated the formation and condensation of fermion pairs in cold atom quantum simulators for dilute extended Hubbard models ($UV$ models) with BCC structure, making calculations of BEC transition temperatures and other pair properties. This is motivated by: (1) a lack of calculations regarding pairing on BCC optical lattices; and (2) BCC superconductors within which fermion pair condensation has been found at high temperatures. To examine pair properties in the dilute limit, we have solved the two-particle Schr\"odinger equation to compute pair mass, radius, transition temperatures and the critical interactions for binding. We performed second-order perturbation theory and QMC calculations to validate our results.  We found that local pairs can form in BCC optical lattices with light masses. We estimated transition temperatures of around 10 nK for Bose-Einstein condensation of pairs for $^6$Li atoms.

The pair mass becomes superlight when $U$ and $V$ are of similar magnitude and both attractive. Then pairs can move freely through the optical lattice without the need to access high-energy intermediate states.  This low mass state coincides with a change from onsite pairs to intersite pairs, and leads to a peak in the transition temperature. Superlight pairs are of particular interest in electron-phonon systems where retarded self-interactions lead to relatively high effective masses even for single particles, and superlight behavior has been predicted to lead to high superconducting transition temperatures \cite{hague2008}. The ability to examine superlight behavior and the properties of pairs in BCC lattices is of interest. We note that pairs form above the condensation temperature, so pair properties are accessible at higher temperatures. 
 Thus, the $UV$ model on BCC lattices offers a system that could be of interest from a condensed matter perspective (particularly from the point of view of superconductivity) and implementable in a clean form by the optical lattices community.

\subsection*{Acknowledgments}
The authors would like to thank Andrey Umerski, Andrew James and Calum MacCormick for useful discussions.

\bibliographystyle{unsrt}
\bibliography{References}

\appendix

\section{Pair Energy, Dispersion and Binding Conditions}

\subsection{Schr\"odinger equation}\label{BCC_singlets_and_triplets}
The (anti-)symmetrized Schr\"odinger equation is given as
\begin{equation}\label{total_WF_appendix}
(E-\varepsilon_{\kvec_{1}}-\varepsilon_{\kvec_{2}})\phi_{\kvec_{1}\kvec_{2}}^{\pm}=\frac{1}{N}\sideset{}{'} \sum_{\qvec\avec_{\pm}}\hat{V}_{\avec_{\pm}}\;\Big\{e^{i(\qvec-\kvec_{1})\,\avec_{\pm}} \pm e^{i(\qvec - \kvec_{2})\,\avec_{\pm}} \Big\}\;\phi_{\qvec,\kvec_{1}+\kvec_{2}-\qvec}^{\pm}
\end{equation}
The prime in the sum implies that a factor of $\frac{1}{2}$ is associated with the case $\avec_{+}=\mathbf{0}$ which takes care of the onsite occupation. The symmetrized Schr\"odinger equation corresponds to spin-singlet states and the anti-symmetrized equation to spin-triplets. For the singlets, we define the vectors
$\{\avec_{+}\} = \{\avec^{+}_{0},\avec^{+}_{1},\avec^{+}_{2},\avec^{+}_{3},\avec^{+}_{4}\}= \{ (0,0,0), (\frac{1}{2},\frac{1}{2},\frac{1}{2}), (-\frac{1}{2},\frac{1}{2},\frac{1}{2}), (\frac{1}{2},-\frac{1}{2},\frac{1}{2}), (\frac{1}{2},\frac{1}{2},-\frac{1}{2})\}\quad$      and for triplets, 
$\{\avec_{-}\} = \{\avec^{-}_{1},\avec^{-}_{2},\avec^{-}_{3},\avec^{-}_{4}\}= \{ (\frac{1}{2},\frac{1}{2},\frac{1}{2}), (-\frac{1}{2},\frac{1}{2},\frac{1}{2}), (\frac{1}{2},-\frac{1}{2},\frac{1}{2}), (\frac{1}{2},\frac{1}{2},-\frac{1}{2})\}$. We set $b=1$ throughout this appendix.

\subsubsection{Symmetrized Sch\"odinger equation}
For the singlets, we use the vectors $\{\avec_{+}\}$ in Equation (\ref{total_WF_appendix}). Thus we have
\begin{equation}\label{bccsingletsWF_appendix}
\begin{split}
& (E - \varepsilon_{\kvec_1} - \varepsilon_{\kvec_2})\phi_{\kvec_{1}\kvec_{2}}^{+} = \frac{1}{N}\sum_{\qvec}\bigg[\frac{1}{2}U(e^{i(\qvec-\kvec_{1})\avec^{+}_{0}}+e^{i(\qvec-\kvec_{2})\avec^{+}_{0}}) + V(e^{i(\qvec-\kvec_{1})\avec^{+}_{1}}+e^{i(\qvec-\kvec_{2})\avec^{+}_{1}})\\
& \:\:\:\:\;\:\:\:\:\; + V(e^{i(\qvec-\kvec_{1})\avec^{+}_{2}}+e^{i(\qvec-\kvec_{2})\avec^{+}_{2}}) + V(e^{i(\qvec-\kvec_{1})\avec^{+}_{3}}+e^{i(\qvec-\kvec_{2})\avec^{+}_{3}})+ V(e^{i(\qvec-\kvec_{1})\avec^{+}_{4}}+e^{i(\qvec-\kvec_{2})\avec^{+}_{4}})\bigg]\phi_{\qvec,\kvec_{1}+\kvec_{2}-\qvec}^{+}\\
& \:\:\:\:\; = \frac{1}{N}\sum_{\qvec}\bigg[U +V\;e^{i(\frac{q_{x}}{2}+\frac{q_{y}}{2}+\frac{q_{z}}{2})}(e^{-i \kvec_{1}\avec^{+}_{1}} + e^{-i \kvec_{2}\avec^{+}_{1}}) +V\; e^{i(\frac{-q_{x}}{2}+\frac{q_{y}}{2}+\frac{q_{z}}{2})}(e^{-i \kvec_{1}\avec^{+}_{2}} + e^{-i \kvec_{2}\avec^{+}_{2}}) \\
& \:\:\:\:\;\;\;\:\:\;\:\:\;\;\;\:\:\;\;\;\:\:\; +V\; e^{i(\frac{q_{x}}{2}-\frac{q_{y}}{2}+\frac{q_{z}}{2})}(e^{-i \kvec_{1}\avec^{+}_{3}} + e^{-i \kvec_{2}\avec^{+}_{3}}) + V\; e^{i(\frac{q_{x}}{2}+\frac{q_{y}}{2}-\frac{q_{z}}{2})}(e^{-i \kvec_{1}\avec^{+}_{4}} + e^{-i \kvec_{2}\avec^{+}_{4}}) \bigg]\phi_{\qvec,\kvec_{1}+\kvec_{2}-\qvec}^{+}
\end{split}
\end{equation}
We can then represent our basis functions as follows:
\begin{equation}\label{bccsingletfinalWF_appendix}
\begin{split}
& \Phi_{0}^{+}(\Pvec) = \frac{1}{N} \sum_{\qvec}\phi_{\qvec,\Pvec-\qvec}^{+} \;\;\;\;\mathrm{,}\;\;\;\;  \Phi_{1}^{+}(\Pvec) = \frac{1}{N} \sum_{\qvec}e^{i(\frac{q_{x}}{2}+\frac{q_{y}}{2}+\frac{q_{z}}{2})}\;\phi_{\qvec,\Pvec-\qvec}^{+} \\
& \Phi_{2}^{+}(\Pvec) = \frac{1}{N} \sum_{\qvec}e^{i(\frac{-q_{x}}{2}+\frac{q_{y}}{2}+\frac{q_{z}}{2})}\;\phi_{\qvec,\Pvec-\qvec}^{+} \;\;\;\mathrm{,}\;\;\; \Phi_{3}^{+}(\Pvec) = \frac{1}{N} \sum_{\qvec}e^{i(\frac{q_{x}}{2}-\frac{q_{y}}{2}+\frac{q_{z}}{2})}\;\phi_{\qvec,\Pvec-\qvec}^{+} \\
& \Phi_{4}^{+}(\Pvec) = \frac{1}{N} \sum_{\qvec}e^{i(\frac{q_{x}}{2}+\frac{q_{y}}{2}-\frac{q_{z}}{2})}\;\phi_{\qvec,\Pvec-\qvec}^{+}
\end{split}
\end{equation}
where $\Pvec = \kvec_{1} + \kvec_{2}$. Hence, Equation (\ref{bccsingletsWF_appendix}) can be written in a more generalized form as
\begin{equation}\label{singlet_all_phi_plus}
\begin{split}
& \phi_{\kvec_{1}\kvec_{2}}^{+} = \frac{1}{(E - \varepsilon_{\kvec_1} - \varepsilon_{\kvec_2})}\bigg\{ U\Phi_{0}^{+}(\Pvec) +V\;\Phi_{1}^{+}(\Pvec)(e^{-i\kvec_{1}\avec^{+}_{1}}+e^{-i\kvec_{2}\avec^{+}_{1}})+V\;\Phi_{2}^{+}(\Pvec)(e^{-i\kvec_{1}\avec^{+}_{2}}+e^{-i\kvec_{2}\avec^{+}_{2}}) \\
& \;\;\;\;\;\;\;\;\;\;\;\;\;\;\;\;+V\;\Phi_{3}^{+}(\Pvec)(e^{-i\kvec_{1}\avec^{+}_{3}}+e^{-i\kvec_{2}\avec^{+}_{3}}) +V\;\Phi_{4}^{+}(\Pvec)(e^{-i\kvec_{1}\avec^{+}_{4}}+e^{-i\kvec_{2}\avec^{+}_{4}})\bigg\}
\end{split}
\end{equation}
We apply Equation (\ref{singlet_all_phi_plus}) to each basis function $\Phi_{0}^{+}(\Pvec)$,$\;\Phi_{1}^{+}(\Pvec)$, $\;\Phi_{2}^{+}(\Pvec)$, $\;\Phi_{3}^{+}(\Pvec)$, $\;\Phi_{4}^{+}(\Pvec)$ and transform the variable $q_{j}$ as:  $q_{j}=q_{j}^{'}+\frac{P_{j}}{2}$. A simple substitution would yield five equations for $\Phi_i^+(\Pvec)$: $i=0,1,2,3,4$ - we give one here as an example.
\begin{equation}
\begin{split}
& \Phi_{0}^{+}(\Pvec) = \frac{1}{N}\sum_{\qvec^{'}}\frac{1}{E-\varepsilon_{\frac{\Pvec}{2}+\qvec^{'}}-\varepsilon_{\frac{\Pvec}{2}-\qvec^{'}}}\bigg\{U\Phi_{0}^{+}(\Pvec) + V\Phi_{1}^{+}(\Pvec)\;e^{-i(\frac{P_{x}}{4}+\frac{P_{y}}{4}+\frac{P_{z}}{4})}\Big[e^{i(\frac{q_{x}^{'}}{2}+\frac{q_{y}^{'}}{2}+\frac{q_{z}^{'}}{2})} + e^{-i(\frac{q_{x}^{'}}{2}+\frac{q_{y}^{'}}{2}+\frac{q_{z}^{'}}{2})}\Big] \\
& \;\;\;\;\;\; + V\Phi_{2}^{+}(\Pvec)\;e^{i(\frac{P_{x}}{4}-\frac{P_{y}}{4}-\frac{P_{z}}{4})}\Big[e^{i(\frac{q_{x}^{'}}{2}-\frac{q_{y}^{'}}{2}-\frac{q_{z}^{'}}{2})} + e^{-i(\frac{q_{x}^{'}}{2}-\frac{q_{y}^{'}}{2}-\frac{q_{z}^{'}}{2})}\Big] + V\Phi_{3}^{+}(\Pvec)\;e^{-i(\frac{P_{x}}{4}-\frac{P_{y}}{4}+\frac{P_{z}}{4})}\Big[e^{i(\frac{q_{x}^{'}}{2}-\frac{q_{y}^{'}}{2}+\frac{q_{z}^{'}}{2})} \\
& \;\;\;\;\;\; + e^{-i(\frac{q_{x}^{'}}{2}-\frac{q_{y}^{'}}{2}+\frac{q_{z}^{'}}{2})}\Big]  + V\Phi_{4}^{+}(\Pvec)\;e^{-i(\frac{P_{x}}{4}+\frac{P_{y}}{4}-\frac{P_{z}}{4})}\Big[e^{i(\frac{q_{x}^{'}}{2}+\frac{q_{y}^{'}}{2}-\frac{q_{z}^{'}}{2})} + e^{-i(\frac{q_{x}^{'}}{2}+\frac{q_{y}^{'}}{2}-\frac{q_{z}^{'}}{2})}\Big] \;\; \bigg\}
\end{split}
\end{equation}
which can be rewritten as
\begin{equation}
    \begin{split}
        & \tilde{\Phi}_{0}^{+}(\Pvec) = UL_{000}(\Pvec)\tilde{\Phi}_{0}^{+}(\Pvec)  + V\Big[L_{111}(\Pvec)+L_{\bar{1}\bar{1}\bar{1}}(\Pvec)\Big]\tilde{\Phi}_{1}^{+}(\Pvec) + V\Big[L_{1\bar{1}\bar{1}}(\Pvec)+L_{\bar{1}11}(\Pvec)\Big]\tilde{\Phi}_{2}^{+}(\Pvec) \\
        & \;\;\;\;\;\;\;\;\;\;\;\;\; + V\Big[L_{1\bar{1}1}(\Pvec)+L_{\bar{1}1\bar{1}}(\Pvec)\Big]\tilde{\Phi}_{3}^{+}(\Pvec) + V\Big[L_{11\bar{1}}(\Pvec)+L_{\bar{1}\bar{1}1)}(\Pvec)\Big]\tilde{\Phi}_{4}^{+}(\Pvec)
    \end{split}
    \label{eqn:phitildezero}
\end{equation}
Note that the new functions,  $\tilde{\Phi}_{i}^{+}(\Pvec)=e^{\frac{-i}{2}(\Pvec\avec_{i}^{+})}\Phi_{i}^{+}$ where $i=0,1,...4$, contain phase factors representing the center-of-mass motion of the pair. In addition, the lattice Green's functions, $L$, is given as 
\begin{align}\label{greens_function_appendix}
    L_{lmn}(\Pvec) & =\frac{1}{N}\sum_{\qvec^{'}}\frac{e^{i(l\frac{q_{x}^{'}}{2}+m\frac{q_{y}^{'}}{2}+n\frac{q_{z}^{'}}{2})}}{E-\varepsilon_{\frac{\Pvec}{2}+\qvec^{'}}-\varepsilon_{\frac{\Pvec}{2}-\qvec^{'}}}=-\int_{-2\pi}^{2\pi}\int_{-2\pi}^{2\pi}\int_{-2\pi}^{2\pi}\frac{dq_{x}^{'} dq_{y}^{'} dq_{z}^{'}}{(4\pi)^{3}}\frac{\cos (l\frac{q_{x}^{'}}{2}+m\frac{q_{y}^{'}}{2}+n\frac{q_{z}^{'}}{2})}{|E|+\varepsilon_{\frac{\Pvec}{2}+\qvec^{'}}+\varepsilon_{\frac{\Pvec}{2}-\qvec^{'}}}
\end{align}
where $l$, $m$ and $n$ are integers $\in$ [0 $\pm$1 $\pm$2]. For compactness, we place a bar above a negative subscript.

For the remaining equations in  (\ref{bccsingletfinalWF_appendix}),
we multiply by $e^{\frac{-i}{2}(\Pvec\avec_{1}^{+})}$, $e^{\frac{-i}{2}(\Pvec\avec_{2}^{+})}$, $e^{\frac{-i}{2}(\Pvec\avec_{3}^{+})}$ and $e^{\frac{-i}{2}(\Pvec\avec_{4}^{+})}$ respectively to obtain equations similar to Eqn (\ref{eqn:phitildezero}).
Hence, self-consistent equations for all spin-singlets at arbitrary momentum can be written as,
\begin{gather}\label{appendix_singlet_sel_cons_eqn}
	\begin{bmatrix}
		UL_{000} & V(L_{111}+L_{\bar{1}\bar{1}\bar{1}}) & V(L_{1\bar{1}\bar{1}}+L_{\bar{1}11} ) & V(L_{1\bar{1}1}+L_{\bar{1}1\bar{1}}) & V(L_{11\bar{1}}+L_{\bar{1}\bar{1}1}) \\
		UL_{111} & V(L_{000}+L_{222}) & V(L_{200} + L_{022}) & V(L_{020} + L_{202}) & V(L_{002} + L_{220}) \\
		UL_{\bar{1}11} & V(L_{\bar{2}00} + L_{022}) & V(L_{000} + L_{\bar{2}22}) & V(L_{\bar{2}20}+L_{002}) & V(L_{\bar{2}02} + L_{020}) \\
		UL_{1\bar{1}1} & V(L_{0\bar{2}0} + L_{202}) & V(L_{2\bar{2}0} + L_{002}) & V(L_{000} + L_{2\bar{2}2}) & V(L_{0\bar{2}2} + L_{200}) \\
		UL_{11\bar{1}} & V(L_{00\bar{2}} + L_{220}) & V(L_{20\bar{2}} + L_{020}) & V(L_{02\bar{2}}+L_{200}) & V(L_{000} + L_{22\bar{2}})
	\end{bmatrix}
	\begin{bmatrix}
	\tilde{\Phi}_{0}^{+}\\ \tilde{\Phi}_{1}^{+}\\ \tilde{\Phi}_{2}^{+}\\ \tilde{\Phi}_{3}^{+}\\ \tilde{\Phi}_{4}^{+}
	\end{bmatrix}
	=
	\begin{bmatrix}
	\tilde{\Phi}_{0}^{+}\\ \tilde{\Phi}_{1}^{+}\\ \tilde{\Phi}_{2}^{+}\\ \tilde{\Phi}_{3}^{+}\\ \tilde{\Phi}_{4}^{+}
	\end{bmatrix}
\end{gather}

\subsubsection{Anti-symmetrized Schr\"odinder equation}
Using $\{\avec_{-}\}$ in Equation (\ref{total_WF_appendix}), the antisymmetrized equation is
\begin{equation}\label{bcctripletsymmetrizedeqution}
\begin{split}
& (E - \varepsilon_{\kvec_1} - \varepsilon_{\kvec_2})\phi_{\kvec_{1}\kvec_{2}}^{-} = \frac{1}{N}\sum_{\qvec}\bigg[ V\;e^{i(\frac{q_{x}}{2}+\frac{q_{y}}{2}+\frac{q_{z}}{2})}(e^{-i \kvec_{1}\avec^{-}_{1}} - e^{-i \kvec_{2}\avec^{-}_{1}}) \\
& \:\:\:\:\;\;\;\:\:\;\:\:\;\;\;\:\:\;\;\;\:\:\; + V\; e^{i(\frac{-q_{x}}{2}+\frac{q_{y}}{2}+\frac{q_{z}}{2})}(e^{-i \kvec_{1}\avec^{-}_{2}} - e^{-i \kvec_{2}\avec^{-}_{2}}) + V\; e^{i(\frac{q_{x}}{2}-\frac{q_{y}}{2}+\frac{q_{z}}{2})}(e^{-i \kvec_{1}\avec^{-}_{3}} - e^{-i \kvec_{2}\avec^{-}_{3}}) \\
& \:\:\:\:\;\;\;\:\:\;\;\;\:\:\;\:\:\:\;\;\;\:\:\;\:\:\;\;\;\:\:\; + V\; e^{i(\frac{q_{x}}{2}+\frac{q_{y}}{2}-\frac{q_{z}}{2})}(e^{-i \kvec_{1}\avec^{-}_{4}} - e^{-i \kvec_{2}\avec^{-}_{4}}) \bigg]\phi_{\qvec,\kvec_{1}+\kvec_{2}-\qvec}^{-}
\end{split}
\end{equation}
Our spin-triplet basis functions are obtained similar to the singlet case as:
\begin{equation}\label{appendix_triplet_phi}
\begin{split}
& \Phi_{1}^{-}(\Pvec) = \frac{1}{N} \sum_{\qvec}e^{i(\frac{q_{x}}{2}+\frac{q_{y}}{2}+\frac{q_{z}}{2})}\;\phi_{\qvec,\Pvec-\qvec}^{-} \;\;\; \mathrm{,} \;\;\; \Phi_{2}^{-}(\Pvec) = \frac{1}{N} \sum_{\qvec}e^{i(\frac{-q_{x}}{2}+\frac{q_{y}}{2}+\frac{q_{z}}{2})}\;\phi_{\qvec,\Pvec-\qvec}^{-} \\
& \Phi_{3}^{-}(\Pvec) = \frac{1}{N} \sum_{\qvec}e^{i(\frac{q_{x}}{2}-\frac{q_{y}}{2}+\frac{q_{z}}{2})}\;\phi_{\qvec,\Pvec-\qvec}^{-} \;\;\; \mathrm{,} \;\;\; \Phi_{4}^{-}(\Pvec) = \frac{1}{N} \sum_{\qvec}e^{i(\frac{q_{x}}{2}+\frac{q_{y}}{2}-\frac{q_{z}}{2})}\;\phi_{\qvec,\Pvec-\qvec}^{-}
\end{split}
\end{equation}

Going through a similar procedure as for the spin-singlets, a set of self-consistent equations for the triplets are obtained. For example,
\begin{equation}
\begin{split}
& \Phi_{1}^{-}(\Pvec) = \frac{1}{N}\sum_{\qvec^{'}}\frac{1}{E-\varepsilon_{\frac{\Pvec}{2}+\qvec^{'}}-\varepsilon_{\frac{\Pvec}{2}-\qvec^{'}}}\bigg\{V\Phi_{1}^{-}(\Pvec)\Big[1 -  e^{2i(\frac{q_{x}^{'}}{2}+\frac{q_{y}^{'}}{2}+\frac{q_{z}^{'}}{2})}\Big] +V\Phi_{2}^{-}(\Pvec)\Big[e^{2i(\frac{q_{x}^{'}}{2})} -  e^{2i(\frac{q_{y}^{'}}{2}+\frac{q_{z}^{'}}{2})}\Big]e^{2i(\frac{P_{x}}{4})} \\
& \;\;\; + V\Phi_{3}^{-}(\Pvec)\Big[e^{2i(\frac{q_{y}^{'}}{2})} -  e^{2i(\frac{q_{x}^{'}}{2}+\frac{q_{z}^{'}}{2})}\Big]e^{2i(\frac{P_{y}}{4})} + V\Phi_{4}^{-}(\Pvec)\Big[e^{2i(\frac{q_{z}^{'}}{2})} - e^{2i(\frac{q_{x}^{'}}{2}+\frac{q_{y}^{'}}{2})}\Big]e^{2i(\frac{P_{z}}{4})} \;\;\; \bigg\} \\
& \;\;\;= V\Big[L_{000}-L_{222}\Big]\Phi_{1}^{-} + Ve^{2i(\frac{P_{x}}{4})}\Big[L_{200}-L_{022}\Big]\Phi_{2}^{-} + Ve^{2i(\frac{P_{y}}{4})} \Big[L_{020)}-L_{202)}\Big]\Phi_{3}^{-} \\
& \;\;\;\;\; + Ve^{2i(\frac{P_{z}}{4})}\Big[L_{002}-L_{220}\Big]\Phi_{4}^{-} 
\end{split}
\end{equation}
This transforms into
\begin{equation}
    \begin{split}
         \\
& \tilde{\Phi}_{1}^{-}(\Pvec)= V\Big[L_{(000}(\Pvec)-L_{222}(\Pvec)\Big]\tilde{\Phi}_{1}^{-}(\Pvec) + V\Big[L_{200}(\Pvec)-L_{022}(\Pvec)\Big]\tilde{\Phi}_{2}^{-}(\Pvec) + V \Big[L_{020}(\Pvec) \\
& \;\;\;\;\; - L_{202}(\Pvec)\Big]\tilde{\Phi}_{3}^{-}(\Pvec) + V\Big[L_{002}(\Pvec)-L_{220}(\Pvec)\Big]\tilde{\Phi}_{4}^{-}(\Pvec)
    \end{split}
\end{equation}
The last step is obtained by multiplying through by the phase factor $e^{\frac{-i}{2}(\Pvec\avec_{1}^{-})}$ such that $\tilde{\Phi}_{1}^{-}(\Pvec)=e^{\frac{-i}{2}(\Pvec\avec_{1}^{-})}\Phi_{1}^{-}$. Likewise, we multiply Equation \ref{appendix_triplet_phi} by, $e^{\frac{-i}{2}(\Pvec\avec_{2}^{-})}$, $e^{\frac{-i}{2}(\Pvec\avec_{3}^{-})}$ and $e^{\frac{-i}{2}(\Pvec\avec_{4}^{-})}$ to obtain the respective expressions for $\tilde{\Phi}_{2}^{-}(\Pvec)$, $\tilde{\Phi}_{3}^{-}(\Pvec)$ and $\tilde{\Phi}_{4}^{-}(\Pvec)$ in (\ref{appendix_triplet_phi}). The expression for the Green's functions, $L$, is the same as defined in Equation (\ref{greens_function_appendix}). Thus, the spin-triplet self-consistent equations at arbitrary momentum can be written as,
\begin{gather}\label{appendix_triplet_sel_cons_eqn}
\begin{bmatrix}
	V(L_{000}-L_{222}) & V(L_{200} - L_{022}) & V(L_{020} - L_{202}) & V(L_{002} - L_{220}) \\
	V(L_{\bar{2}00} - L_{022}) & V(L_{000} - L_{\bar{2}22}) & V(L_{\bar{2}20}-L_{002}) & V(L_{\bar{2}02} - L_{020}) \\
	V(L_{0\bar{2}0} - L_{202}) & V(L_{2\bar{2}0} - L_{002}) & V(L_{000} - L_{2\bar{2}2}) & V(L_{0\bar{2}2} - L_{200}) \\
	V(L_{00\bar{2}} - L_{220}) & V(L_{20\bar{2}} - L_{020}) & V(L_{02\bar{2}}-L_{200}) & V(L_{000} - L_{22\bar{2}})
	\end{bmatrix}
\begin{bmatrix}
	\tilde{\Phi}_{1}^{-}\\ \tilde{\Phi}_{2}^{-}\\ \tilde{\Phi}_{3}^{-}\\ \tilde{\Phi}_{4}^{-}
	\end{bmatrix}
=
\begin{bmatrix}
	\tilde{\Phi}_{1}^{-}\\ \tilde{\Phi}_{2}^{-}\\ \tilde{\Phi}_{3}^{-}\\ \tilde{\Phi}_{4}^{-}
	\end{bmatrix}
\end{gather}
where $L$ is the Green's function defined earlier in (\ref{greens_function_appendix}).

Equations (\ref{appendix_singlet_sel_cons_eqn}) and (\ref{appendix_triplet_sel_cons_eqn}) are both eigenequations that must be solved to obtain pair properties. 

\subsection{Pair energy for $\Gamma$ point}
Next, we compute the energies of the singlets and triplets. At the $\Gamma$ point where ($P_x=P_y=P_z=0$) - there is a further simplification of the Green's functions (\ref{greens_function_appendix}) which can be expressed as an integral, 
\begin{equation}
\begin{split}
& L_{lmn}(0)=\frac{1}{N}\sum_{\qvec^{'}}\frac{e^{i(l\frac{q_{x}^{'}}{2}+m\frac{q_{y}^{'}}{2}+n\frac{q_{z}^{'}}{2})}}{E-2\varepsilon_{\qvec^{'}}}=-\int_{-2\pi}^{2\pi}\int_{-2\pi}^{2\pi}\int_{-2\pi}^{2\pi}\frac{dq_{x}^{'} dq_{y}^{'} dq_{z}^{'}}{(4\pi)^{3}}\frac{\cos (l\frac{q_{x}^{'}}{2}+m\frac{q_{y}^{'}}{2}+n\frac{q_{z}^{'}}{2})}{|E|- 16t\cos(\frac{q_{x}^{'}}{2})\cos(\frac{q_{y}^{'}}{2})\cos(\frac{q_{z}^{'}}{2})} \\ & =-\frac{1}{(2\pi)^3} \int_{-\pi}^{\pi}\int_{-\pi}^{\pi}\int_{-\pi}^{\pi}\frac{\cos(lq_{x}^{''})\cdot\cos(mq_{y}^{''})\cdot\cos(nq_{z}^{''})}{|E|-16t\cos (q_{x}^{''})\cdot\cos(q_{y}^{''})\cdot\cos(q_{z}^{''})}dq_{x}^{''}\,dq_{y}^{''}dq_{z}^{''}\;\;: \;\;\;(q_{j}^{''} = \frac{q_{j}^{'}}{2})
\end{split}
\end{equation}
and the following relations hold
\begin{equation}\label{all_gf_at_gamma_point}
\begin{split}
	& L_{000}(0) \equiv L_{0} \\
	& L_{111}(0)=L_{\bar{1}\bar{1}\bar{1}}(0)=L_{\bar{1}11}(0)=L_{1\bar{1}\bar{1}}(0)=L_{1\bar{1}1}(0)=L_{\bar{1}1\bar{1}}(0)=L_{11\bar{1}}(0)=L_{\bar{1}\bar{1}1}(0) \equiv L_{1} \\
	& L_{222}(0)=L_{\bar{2}22}(0)=L_{2\bar{2}2}(0)=L_{22\bar{2}}(0) \equiv L_{2} \\
	& L_{200}(0)=L_{\bar{2}00}(0)=L_{020}(0)=L_{0\bar{2}0}(0)=L_{002}(0)=L_{00\bar{2}}(0) \equiv L_{3} \\
	& L_{022}(0)=L_{0\bar{2}2}(0)=L_{02\bar{2}}(0)=L_{202}(0)=L_{20\bar{2}}(0)=L_{\bar{2}02}(0)=L_{220}(0)=L_{2\bar{2}0}(0)=L_{\bar{2}20}(0) \equiv L_{4}
\end{split}
\end{equation}
Then (\ref{appendix_singlet_sel_cons_eqn}) and (\ref{appendix_triplet_sel_cons_eqn}) respectively becomes (Note that $\tilde{\Phi}_{i}^{\pm}\equiv\Phi_{i}^{\pm}$ since $\Pvec=0$) 
\begin{gather}
	\underbrace{\begin{bmatrix}
		UL_{0} & 2VL_{1} & 2VL_{1} & 2VL_{1} & 2VL_{1} \\
		UL_{1} & V(L_{0} + L_{2}) & V(L_{3} + L_{4}) & V(L_{3} + L_{4}) & V(L_{3} + L_{4}) \\
		UL_{1} & V(L_{3} + L_{4}) & V(L_{0} + L_{2}) & V(L_{4}+L_{3}) & V(L_{4} + L_{3}) \\
		UL_{1} & V(L_{3} + L_{4}) & V(L_{4} + L_{3}) & V(L_{0}+L_{2}) & V(L_{4} + L_{3}) \\
		UL_{1} & V(L_{3} + L_{4}) & V(L_{4} + L_{3}) & V(L_{4}+L_{3}) & V(L_{0} + L_{2})
	\end{bmatrix}}_{\hat{L}_{\rm singlet}}
	\underbrace{\begin{bmatrix}
	\Phi_{0}^{+}\\ \Phi_{1}^{+}\\ \Phi_{2}^{+}\\ \Phi_{3}^{+}\\ \Phi_{4}^{+}
	\end{bmatrix}}_{\hat{\Phi}_{\rm singlet}}
	=
	\underbrace{\begin{bmatrix}
	\Phi_{0}^{+}\\ \Phi_{1}^{+}\\ \Phi_{2}^{+}\\ \Phi_{3}^{+}\\ \Phi_{4}^{+}
	\end{bmatrix}}_{\hat{\Phi}_{\rm singlet}}
\end{gather}
\begin{gather}
\underbrace{\begin{bmatrix}
	V(L_{0} - L_{2}) & V(L_{3} - L_{4}) & V(L_{3} - L_{4}) & V(L_{3} - L_{4}) \\
	V(L_{3} - L_{4}) & V(L_{0} - L_{2}) & V(L_{4} - L_{3}) & V(L_{4} - L_{3}) \\
	V(L_{3} - L_{4}) & V(L_{4} - L_{3}) & V(L_{0} - L_{2}) & V(L_{4} - L_{3}) \\
	V(L_{3} - L_{4}) & V(L_{4} - L_{3}) & V(L_{4} - L_{3}) & V(L_{0} - L_{2})
	\end{bmatrix}}_{\hat{L}_{\rm triplet}}
\underbrace{\begin{bmatrix}
	\Phi_{1}^{-}\\ \Phi_{2}^{-}\\ \Phi_{3}^{-}\\ \Phi_{4}^{-}
	\end{bmatrix}}_{\hat{\Phi}_{\rm triplet}}
=
\underbrace{\begin{bmatrix}
	\Phi_{1}^{-}\\ \Phi_{2}^{-}\\ \Phi_{3}^{-}\\ \Phi_{4}^{-}
	\end{bmatrix}}_{\hat{\Phi}_{\rm triplet}}
\end{gather}
The matrix equations above can be written in a compact form as
\begin{equation}
	\hat{L}_{s,t}\,\hat{\Phi}_{s,t}= \lambda_{s,t}\,\hat{\Phi}_{s,t}
\end{equation}
thus forming an eigenvalue problem. $\hat{L}_{s}$ and $\hat{L}_{t}$ are the singlet and triplet dispersion matrices, $\lambda_{s}$ and $\lambda_{t}$ being the eigenvalues corresponding to singlet $\hat{\Phi}_{s}$ and triplet $\hat{\Phi}_{t}$ eigenvectors respectively.
So, the values of $E$ corresponding to the pair energy are found by searching for eigenvalues of $\lambda_{s,t}=1$. 

\subsection{Further symmetrization at the Brillouin zone center}\label{appendix:symmetrization_at_the_gamma_point}

The system of equations can be significantly simplified at the $\Gamma$ point of the BCC BZ, where the system possesses $O_{h}$ point symmetry, and this will enable the calculation of binding criteria for pairs at the $\Gamma$ point with $s$, $p$, $d$, and $f$ symmetry. By performing the 48 operations on the system, we determine the linear combinations (excluding the normalization constants) of the eigenvector through the irreducible representations of the $O_{h}$ group\cite{JFCornwell1997}. The eigenfunction $\Phi_{0}^{+}$ is located at the center of the BZ and it remains unchanged as all operations are performed (thus forming an \textsl{s}-like molecular orbital).
The sought irreducible representation for both the singlet and triplet states are
\begin{equation}
\begin{split}
 & \Gamma_{\rm singlet}^{\rm bcc} = A_{1g} \oplus T_{2g} \\
 & \Gamma_{\rm triplet}^{\rm bcc} = T_{1u} \oplus A_{2u} 
 \end{split}
 \end{equation}
 whence $A_{1g}$, $T_{2g}$, $A_{2u}$ and $T_{2u}$ forms the $s$-, $d$-, $p$- and $f$- states respectively. From here, we can find the symmetrized linear combinations for the singlets, $\Gamma_{\rm singlet}^{\rm bcc}$, as 
 \begin{align}\label{bcc_singlets_linear_combination}
	 \chi_{}^{A_{1g}} & = \Phi_{1}^{+} + \Phi_{2}^{+} + \Phi_{3}^{+} + \Phi_{4}^{+} \\
	 \chi_{}^{T_{2g}} & = 
		\begin{cases}
			\Phi_{1}^{+} + \Phi_{2}^{+} - \Phi_{3}^{+} - \Phi_{4}^{+} \\
			\Phi_{1}^{+} - \Phi_{2}^{+} \\
			\Phi_{3}^{+} - \Phi_{4}^{+} \\
	\end{cases}
 \end{align}

\noindent and the triplets, $\Gamma_{\rm triplet}^{\rm bcc}$, have the combinations 
 \begin{align}
  \chi_{}^{T_{1u}} & = 
 \begin{cases}
 \Phi_{1}^{-} - \Phi_{2}^{-} + \Phi_{3}^{-} + \Phi_{4}^{-} \\
 \Phi_{1}^{-} + \Phi_{2}^{-} \\
 \Phi_{3}^{-} - \Phi_{4}^{-} \\
 \end{cases} \\
 \chi_{}^{A_{2u}} & = \Phi_{1}^{-} - \Phi_{2}^{-} - \Phi_{3}^{-} - \Phi_{4}^{-}
 \end{align}
If we combine Equation(\ref{bcc_singlets_linear_combination}) with $\Phi_{0}^{+}$, we get the transformation to new basis \footnote{Remember that we have omitted the normalization factors of the new symmetrized basis.}$^,$\footnote{The subscript $s$ has been used twice: $\hat{\Phi}_{s}$ means all possible singlet states ($s,d,\dots$) while $\Phi_{s}$ means an $s$-state.}
\begin{gather}
\hat{\Phi}_{s}
=
\begin{bmatrix}
\Phi_{0}\\ \Phi_{s}\\ \Phi_{d_{1}} \\ \Phi_{d_{2}} \\ \Phi_{d_{3}}
\end{bmatrix}
=
\begin{bmatrix}
1 & 0 & 0 & 0 & 0\\
0 & 1 & 1 & 1 & 1 \\
0 & 1 & 1 & -1 & -1\\
0 & 1 & -1 & 0 & 0\\
0 & 0 & 0 & 1 & -1
\end{bmatrix}
\begin{bmatrix}
\Phi_{0}^{+}\\
\Phi_{1}^{+}\\
\Phi_{2}^{+}\\
\Phi_{3}^{+}\\
\Phi_{4}^{+}
\end{bmatrix}
\equiv
\hat{\chi_{s}}
\begin{bmatrix}
\Phi_{0}^{+}\\
\Phi_{1}^{+}\\
\Phi_{2}^{+}\\
\Phi_{3}^{+}\\
\Phi_{4}^{+}
\end{bmatrix}
\end{gather}
\begin{gather}
\hat{\Phi}_{t}
=
\begin{bmatrix}
\Phi_{p_{1}}\\ \Phi_{p_{2}} \\ \Phi_{p_{3}} \\ \Phi_{f}
\end{bmatrix}
=
\begin{bmatrix}
1 & -1 & 1 & 1 \\
1 & 1 & 0 & 0\\
0 & 0 & 1 & -1\\
1 & -1 & -1 & -1
\end{bmatrix}
\begin{bmatrix}
\Phi_{1}^{-}\\
\Phi_{2}^{-}\\
\Phi_{3}^{-}\\
\Phi_{4}^{-}
\end{bmatrix}
\equiv
\hat{\chi_{t}}
\begin{bmatrix}
\Phi_{1}^{-}\\
\Phi_{2}^{-}\\
\Phi_{3}^{-}\\
\Phi_{4}^{-}
\end{bmatrix}
\end{gather}

\noindent Actually, $\hat{\chi}_{i}$ (where $i = s,t$) is derived by performing the symmetry operations and it can help diagonalize the problem further via the equation
\begin{equation}
	\hat{L}_{i}^{\rm diag}=\hat{\chi}_{i}\cdot \hat{L}_{i}\cdot \hat{\chi}_{i}^{-1}
\end{equation}
By applying the formula above, the new symmetrized bases $\hat{\chi}_{s}$ and $\hat{\chi}_{t}$ respectively block-diagonalizes the dispersion relations $\hat{L}_{s}$ and $\hat{L}_{t}$ as follows  

\begin{gather}\label{bcc_diag_matrix_singlet}
\begin{bmatrix}
UL_{0} & 2VL_{1} & 0 & 0 & 0 \\
4UL_{1} & \mathcal{K}_{s} & 0 & 0 & 0 \\
0 & 0 & \mathcal{K}_{d} & 0 & 0 \\
0 & 0 & 0 & \mathcal{K}_{d} & 0 \\
0 & 0 & 0 & 0 & \mathcal{K}_{d}
\end{bmatrix}
\begin{bmatrix}
\Phi_{0}\\ \Phi_{s}\\ \Phi_{d_{1}}\\ \Phi_{d_{2}}\\ \Phi_{d_{3}}
\end{bmatrix}
=
\begin{bmatrix}
\Phi_{0}\\ \Phi_{s}\\ \Phi_{d_{1}}\\ \Phi_{d_{2}}\\ \Phi_{d_{3}}
\end{bmatrix}
\end{gather}

\begin{gather}\label{bcc_diag_matrix_triplet}
\begin{bmatrix}
\mathcal{K}_{p} & 0 & 0 & 0 \\
0 & \mathcal{K}_{p} & 0 & 0 \\
0 & 0 & \mathcal{K}_{p} & 0 \\
0 & 0 & 0 & \mathcal{K}_{f}
\end{bmatrix}
\begin{bmatrix}
\Phi_{p_{1}}\\ \Phi_{p_{2}} \\ \Phi_{p_{3}} \\ \Phi_{f}
\end{bmatrix}
=
\begin{bmatrix}
\Phi_{p_{1}}\\ \Phi_{p_{2}} \\ \Phi_{p_{3}} \\ \Phi_{f}
\end{bmatrix}
\end{gather}

\noindent $\mathcal{K}_{s}=V(L_{0} + L_{2} + 3L_{3}+3L_{4})$, $\mathcal{K}_{d}=V(L_{0} + L_{2} - L_{3} - L_{4})$, $\mathcal{K}_{p}=V(L_{0} - L_{2} + L_{3} - L_{4})$, $\mathcal{K}_{f}=V(L_{0} - L_{2} - 3L_{3} + 3L_{4})$.\\

In (\ref{bcc_diag_matrix_singlet}), the top-left $2\times2$ block corresponds to the $s$-symmetrical state while the other three $1\times1$ blocks are $d$-symmetrical states which are triply degenerate (when $\Pvec=0$).
Similarly in Equation (\ref{bcc_diag_matrix_triplet}), the $p$-states are 3-fold degenerate and there is a single $f$-state. Note that this diagonalization is valid for the entire $\Gamma$-H line where the symmetry at the $\Gamma$ point is retained.

\subsection{Binding criterion at the Brillouin zone center}

Once the system of equations has been symmetrized, it is possible to calculate the critical binding threshold by setting $E\rightarrow-2W=-16t$. For this purpose, it is convenient to re-write the singlet and triplet determinant matrices (\ref{bcc_diag_matrix_singlet}) and (\ref{bcc_diag_matrix_triplet}), and remove redundant elements\\

\setlength{\belowdisplayskip}{0pt} \setlength{\belowdisplayshortskip}{0pt}
\setlength{\abovedisplayskip}{0pt} \setlength{\abovedisplayshortskip}{0pt}
\begin{align}\label{s_critical_matrix}
s:\hspace{7ex}\begin{bmatrix}
1 - UL_{0} & -2VL_1 \\
-4UL_{1} & 1-\mathcal{K}_{s}
\end{bmatrix}=0
\end{align}
\begin{align}\label{critical_v_nonS}
     d:\hspace{13.3ex} &1-\mathcal{K}_{d} =0 \hspace{7ex}\\
     p:\hspace{13.3ex} &1 - \mathcal{K}_{p}= 0 \\
     f:\hspace{13.3ex} &1 - \mathcal{K}_{f}=0 \label{critical_vf}
\end{align}\medskip

The Green's functions (\ref{all_gf_at_gamma_point}) can be expressed in terms of the elliptic integral of the first kind\cite{joyce1971exacintegraltbccgreenfunction} as \\
\begin{align}
    L_{0}&=-\frac{K_{0}^2}{4\pi^2t} =\frac{-0.087075245605354804}{t} \\
    L_{1}&=\frac{1}{16t} -\frac{K_{0}^2}{4\pi^2t} =\frac{1}{16t} + L_{0} \\
    L_{2}&=\frac{1}{2t} -\frac{K_{0}^2}{\pi^2t} -\frac{9}{16tK_{0}^2}=\frac{1}{2t} + 4L_{0} + \frac{9}{64\pi^2t^2L_{0}} \\
    L_{3}&=-\frac{1}{16tK_{0}^2}=\frac{1}{64\pi^2t^2L_{0}}\\
    L_{4}&=-\frac{K_{0}^2}{4\pi^2t} + \frac{1}{4tK_{0}^2}=L_{0}-\frac{1}{16\pi^2t^2L_0}=L_0 -4L_3
\end{align}
where $K_{0}=K\big(\frac{1}{\sqrt{2}}\big)=1.85407467\dots$ is the complete elliptic integral of the first kind.\medskip

Expanding  the determinant (\ref{s_critical_matrix})  gives the critical binding expression \\
\begin{equation}
    V_{c}^{s}\leq V(U)=\frac{UL_{0}-1}{UL_{0}\mathcal{C}-\mathcal{C}-8UL_{1}^2}
\end{equation}\\
where $\mathcal{C}=L_{0} + L_{2} + 3L_{3}+3L_{4}=8L_0 + \frac{1}{2t}=-0.19660196484283837/t$.\\
Therefore,
\begin{align}
    &V_{c}^{s}(U=0) =-5.0864191t\\
    &V_{c}^{s}(U\rightarrow+\infty)=-7.0864191t \\
    &U_{c}(V=0) =-11.4843202t \\
    &U_{c}(V\rightarrow+\infty) =-16t
\end{align}

Similarly, (\ref{critical_v_nonS}) - (\ref{critical_vf}) respectively yields
\begin{align}
    V_c^d=-15.0428185t\\
    V_c^p=-12.6624416t \\
    V_c^f=-15.7113739t
\end{align}

\end{document}